\documentclass[twocolumn,superscriptaddress,aps,prl,amssymb,amsfonts,amsmath]{revtex4-1}
\usepackage{bm}
\usepackage{graphicx}
\usepackage{xcolor}

\definecolor{bleuf}{rgb}{0,0.44,0.72}
\usepackage[unicode=true,pdfusetitle,bookmarks=true,
bookmarksnumbered=false,bookmarksopen=false,breaklinks=false,
pdfborder={0 0 0},backref=false,colorlinks=true,citecolor=bleuf,
linkcolor=bleuf,urlcolor=bleuf]{hyperref}

\usepackage{caption}
\renewcommand\vec{\mathbf}
\newcommand{\bb}{\vec{b}}

\newcommand{\FF}{\vec{F}}

\newcommand{\rr}{\vec{r}}
\newcommand{\RR}{\vec{R}}

\begin{document}

\title{Melting of non reciprocal solids: how dislocations propel and fission in flowing crystals}

\author{St\'ephane Guillet}
\author{Alexis Poncet}
\affiliation{ENS de Lyon, CNRS, LPENSL, UMR5672, 69342, Lyon cedex 07, France}
\author{Marine Le Blay}
\affiliation{Pôle d’Etudes et Recherche de Lacq, Total SA, BP 47-64170 Lacq, France}
\affiliation{ENS de Lyon, CNRS, LPENSL, UMR5672, 69342, Lyon cedex 07, France}
\author{William. T. M Irvine}
\author{Vincenzo Vitelli}
\affiliation{James Franck Institute, University of Chicago, Chicago, IL 60637}
\affiliation{Department of Physics, University of Chicago, Chicago, IL 60637}
\author{Denis Bartolo}
\affiliation{ENS de Lyon, CNRS, LPENSL, UMR5672, 69342, Lyon cedex 07, France}

\date{\today}

\begin{abstract}
 When soft matter is driven out of equilibrium its constituents  interact via effective interactions that escape Newton's action-reaction principle. 
Prominent examples include the hydrodynamic interactions between colloidal particles driven in viscous fluids, phoretic interactions between chemically active colloids, and  quorum sensing interactions in bacterial colonies. 
Despite a recent surge of interest in non-reciprocal physics 
a fundamental question remains: do non-reciprocal interactions alter or strengthen the ordered phases of matter driven out of equilibrium? 
Here, through a combination of  experiments and simulations, we show how nonreciprocal  forces propel and fission dislocations formed in hydrodynamically driven Wigner crystals.
We explain how dislocation motility results in the continuous reshaping of  grain-boundary networks,  and how their fission reaction melts driven crystals from their interfaces.
Beyond the specifics of hydrodynamics, we argue theoretically  that topological defects and nonreciprocal interactions should invariably conspire to deform and ultimately destroy crystals whose the elementary units defy Newton's third law
\end{abstract}

\maketitle

In our daily experiences, we accurately perceive Newton’s laws of mechanics: we do not feel any force as we travel at constant speed in a train, we  feel a force when it accelerates, and  we feel the back-reaction of our seat on our back as we compress it with our body weight.  
However, our intuition  is significantly challenged when trying to understand the behavior of 
systems composed of many interacting bodies, for which we
only have access to a small subset of  their degrees of freedom. 
Consider the simple example of two rigid particles falling in a viscous fluid under the sole action of gravity.  
It is  surprising to observe that their speed and trajectory depend on their relative positions~\cite{happel2012low,Chawa2019}. 
Two beads making a finite angle with the vertical axis do not even fall straight. 
The reason for this counter intuitive behavior is that our observations ignore the many degrees of freedom of the surrounding fluid. 
As a bead falls, it couples to the fluid flow, which in turn induces a drag force on the second particle. 
This dynamical coupling is commonly referred to as a hydrodynamic interaction~\cite{happel2012low,Brady1988}. 
However, unlike all conservative forces that derive from an energy potential, hydrodynamic interactions do not obey Newton's action-reaction principle.
The drag forces induced by the two particles on one another are not opposite: hydrodynamic interactions are non-reciprocal.

Beyond hydrodynamics, non-reciprocal  interactions rule the dynamics of systems as diverse as chemically active colloids~\cite{Palacci2013,Soto_2014}, coupled robots~\cite{Fruchart_2021,Veenstra2024}, and interacting living creatures, from bacteria~\cite{seymour2023}, to birds~\cite{Cavagna2014} and human beings~\cite{Karamouzas2014}.  

The impact of non-reciprocal forces on many-body physics offers some formidable problems in fluid mechanics, soft condensed matter and statistical mechanics.
Even in the paradigmatic situation where a collection of underdamped particles are uniformly driven in a viscous fluid, their trajectories do not form mere straight lines, they are chaotic and strongly correlated in space and time~\cite{Guazzelli2011,Ramaswamy2001,Beatus2017}. 
After decades of intense studies, the anomalous statistics of the velocity fluctuations, and  the structural arrangements of hydrodynamically coupled bodies are yet to be elucidated and cannot be captured by an effective (Maxwell-)Boltzman statistics. 
Even when they are arranged on a perfect periodic lattice, the  dynamics of particles driven in viscous fluids remains highly counter intuitive.
In the absence of any stabilizing elastic forces, theory predicts that fluctuations in the particle positions can cause the propagation of singular sound waves  despite the irrelevance of inertia, or be amplified to destroy their spatial ordering~\cite{Crowley71,Beatus_2006,Desreumaux_2012,Chawa2020,Saeed2023}. 

In this article, our objective is to answer a fundamental question which remains unsolved despite a growing interest in non-reciprocal matter, see e.g.~\cite{Soto2014,Ivlev2015,Meredith2020,You2020,Saha2020,Fruchart_2021b,Osat2023,Golestanian2022,Poncet2022,Maity2023,Dinelli2023,Chen2024,Rouzaire2024}: How do fluctuations powered by non-reciprocal forces alter the ordered phases of soft matter? 
To make progress, we consider a basic situation where crystalline order is stabilized by conservative forces and undermined by hydrodynamic interactions.
We establish two universal results through the combination of experiments, numerical simulations and mathematical modeling: 
(i) Non-reciprocal  interactions sustain the continuous reshaping of the grain boundaries that separate crystals of incompatible orientations.
(ii) When non-reciprocal forces overcome elastic interactions, they split dislocations into pairs and cause their exponential proliferation. 
This nonequilibrium 2D melting process escapes the standard KTHNY~\cite{Halperin1978,Nelson_1979,Young1979} and hydrodynamic clumping scenarios~\cite{Crowley71,Ramaswamy2001,Saeed2023}.

\begin{figure*}[ht]
        \centering
        \includegraphics[width=0.9\textwidth]{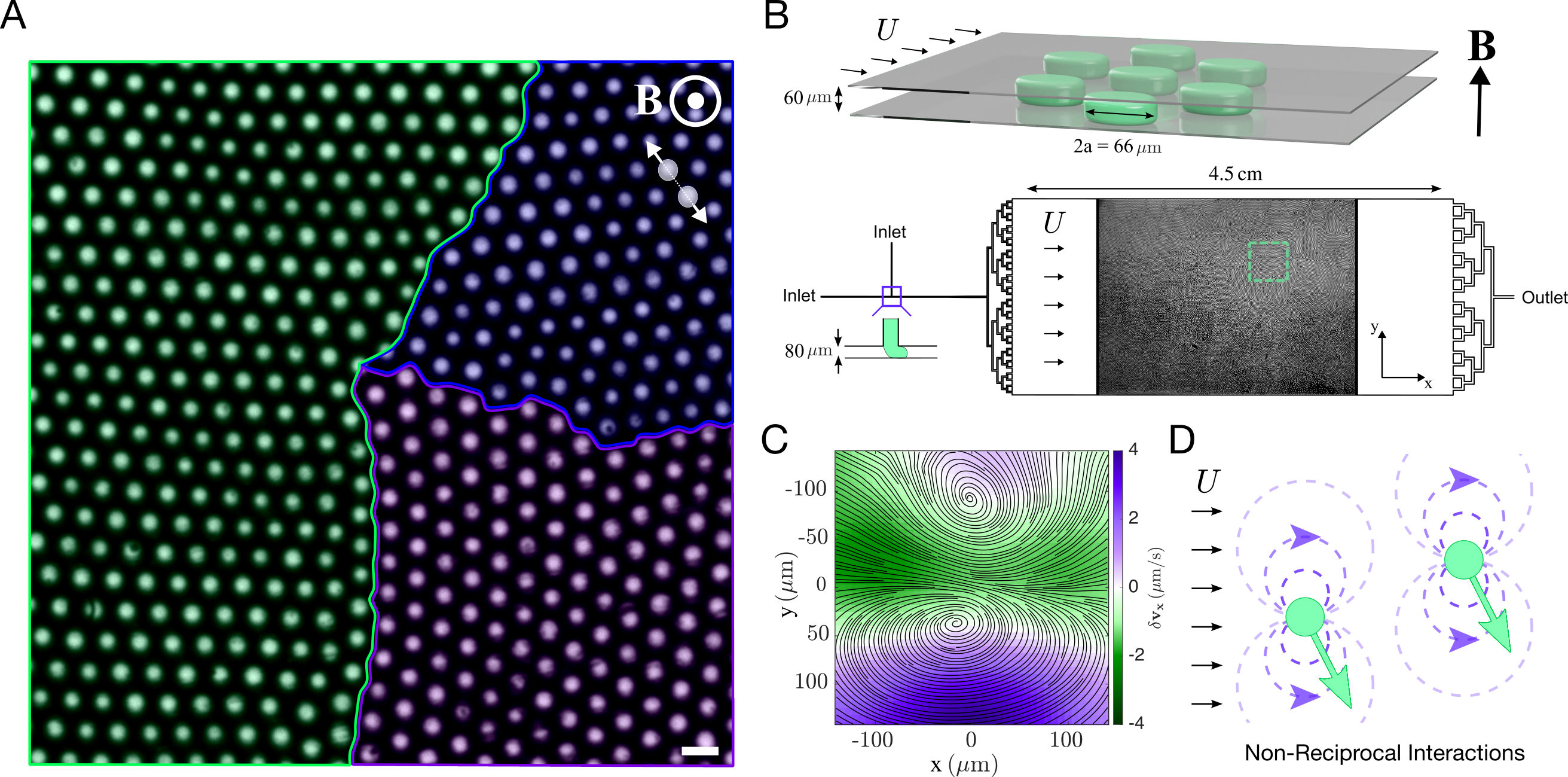}
        \caption{\textbf{Experimental system: hydrodynamically driven Wigner crystals.}
        {\bf A.} Experimental image of a two-dimensional collection of hexadecane droplets dispersed in an aqueous ferrofluid.
        When applying a magnetic field, the droplets repel one another (white arrows), and self organize into   Wigner crystals separated by grain boundaries.
        The color shade indicates the orientation of the principal axis of the crystallites.
        The wite arrows indicate that the magnetic repulsion between the dipoles obey Newton's action reaction principle.
        Scale bar : $150\rm{\,\mu m}$. 
        {\bf B.} Schematics of our microfluidic device.
        We continuously produce a monodisperse emulsion at a T-junction and fill a $4.5\,\rm cm$ long Hele-Shaw cell.
        The oil droplets
         are confined by the two vertical walls and have a pancake shape. 
        The device is placed in an electric coil, the resulting magnetic field $\mathbf B$ points in the direction transverse to the confining walls and induce  repulsive interactions between the drops. 
        The resulting polycrystal is advected at constant flow rate by a homogeneous plug flow $\mathbf U$ that points in the $x$ direction (horizontal arrows).
        {\bf C.} Streamlines of the velocity field constructed from the droplets' motion. 
        To quantify the hydrodynamic interactions,
        we track a droplet and use it to define a  reference frame. 
        We measure the velocity of all the surrounding droplets in this moving frame. We then construct the corresponding velocity field and time average it. We also average the droplet-velocity field over a hundreds of different reference droplets. We can clearly see that the dominant contribution to the hydrodynamic interactions has a dipolar symmetry. The color indicates the magnitudes of the longitudinal velocity in the moving frame.
        {\bf D.} Sketch of the far-field hydrodynamic force field around two droplets. 
        The hydrodynamic interactions do not obey Newton's third law: they are non-reciprocal.}
        \label{fig:fig1}
    \end{figure*} 

We organize the presentation of our results as follows. We start with an introduction of our experiments based on driven crystalline emulsions, and our  numerical model, providing a bird’s-eye view of our main findings. 
Next, we thoroughly characterize the dislocation dynamics and melting of driven polycrystals in response to nonreciprocal hydrodynamic interactions. Finally, we conclude with a general discussion.
Beyond the specifics of our experiments, we explain how the interplay between dislocation dynamics and nonreciprocal forces governs the behavior of a broad class of nonequilibrium crystals, where conservative  forces compete with nonreciprocal interactions arising from hydrodynamic, mechanical, chemical, or cognitive couplings.

\section*{Driving Wigner crystals out of equilibrium with fluid flows: Experiments, simulations and main findings.}

    \subsection*{Experiments}
    We study the dynamics and structure of hydrodynamically-driven crystals through the experiment shown in Figure~\ref{fig:fig1} and detailed in SI. In short,
    we assemble 2D Wigner crystals, by applying a uniform magnetic field to a paramagnetic emulsion confined in a microfluidic Hele-Shaw cell sketched in Figure~\ref{fig:fig1}B. 
    The resulting magnetic forces between the droplets (radius $33\,\rm \mu m$) are isotropic, repulsive and decay as $1/r^4$, Figure~\ref{fig:fig1}A.
    They classically lead to the formation of pristine crystallites of incompatible orientations separated by grain boundaries,  Figure~\ref{fig:fig1}A.
    Variations of this standard setup have been instrumental to shed light on the statistics and kinetics of thermal melting in two dimensions~\cite{Zanh_1999,Grunberg2007,Gasser2010,Deutschlander_2015}.
    Here, we drive the system out of equilibrium with a pressure gradient transverse to the magnetic field. 
    In the absence of droplets, the pressure drop results in a uniform flow of the continuous phase along the $x$ direction, Figure~\ref{fig:fig1}B. 
    However the droplets do not behave as passive tracers.
    Due to  viscous friction on the confining walls, they are advected at a speed that is about two times smaller that the average flow~\cite{Desreumaux_2013}.
    This velocity mismatch  is key to inducing non-reciprocal interactions.
    Each droplet deforms the otherwise uniform flow and yield dipolar recirculations that decay algebraically in space as $1/r^2$, see e.g~\cite{Beatus_2006}.
    The droplets are then advected both by the mean flow and the dipolar disturbances induced by the surrounding particles. 
    To demonstrate the relevance of these hydrodynamic interactions, we measure the time-averaged displacement of the droplets in a reference frame moving with a tagged particle. Figure~\ref{fig:fig1}C shows that, when $B=0\,\rm  mT$, the particle trajectories are clearly perturbed by dipolar recirculations (see also SI).
    Crucially, as illustrated in Figure~\ref{fig:fig1}D, when two particles are advected by these hydrodynamic perturbations, the resulting drag forces have equal magnitudes but not opposite directions. Unlike magnetic forces, hydrodynamic forces do not obey Newton’s action-reaction principle. They are non-reciprocal.
    
    We study the competition between the reciprocal (magnetic) and non-reciprocal (hydrodynamic) interactions by tuning their relative magnitude.
    We keep the mean-flow speed constant (corresponding to a mean droplet speed of $U=150\pm1\,\rm \mu m/s$) and vary the strength of the $\mathbf B$ field between $B=0\,\rm mT$ to $B=10\,\rm  mT$ using a $50\,\rm cm$ cylindrical coil.
    All the results presented in the main text correspond to experiments where the packing fraction is set to $\Phi=0.37$ in the main channel.  We keep the packing fraction constant throughout the experiments, by continuously injecting droplets in the main channel as we drive the emulsions. 
    In SI, we provide a comprehensive description of our experimental setup and report a second series of experiments where $\Phi=0.61$. 
    None of our observations and conclusions crucially depend on these values.

    \begin{figure*}[ht]
        \centering
        \includegraphics[width=\textwidth]{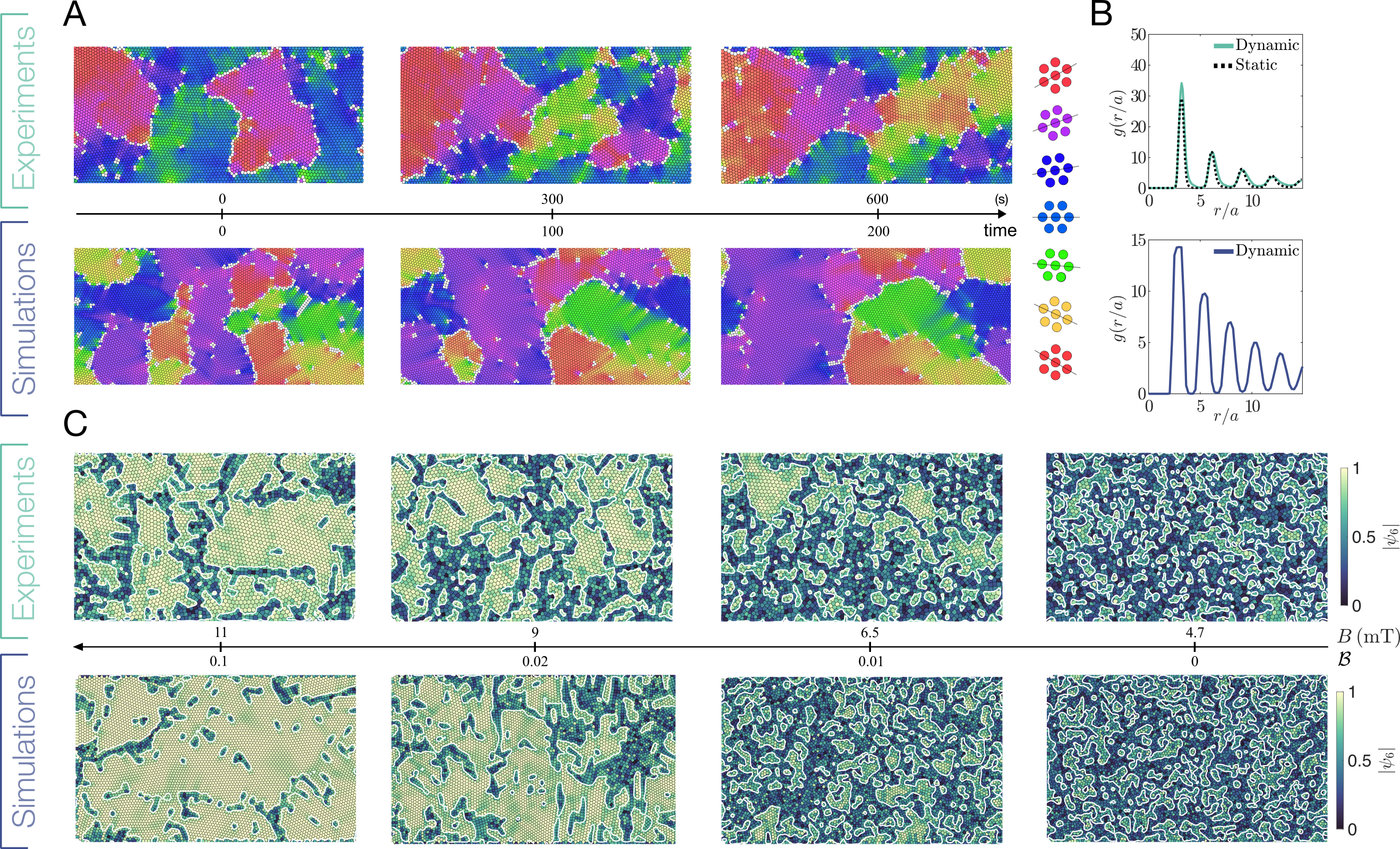}  
        \caption{\textbf{Polycrystals dynamic : experiments vs numerics.}
        {\bf A.} Four subsequent snapshots of the crystal-orientation field defined as the phase $\theta_6$ of the  hexatic order parameter $\psi_6$~\cite{kardar2007fields}.
        Top row: experiments. Bottom row: numerical simulations.
        Both our experiments ($B=9.7\,\rm mT$) and simulations ($\mathcal B=0.2$) reveal that the driven polycrystals continuously reshape the geometry of their grain boundaries.
        The $\theta_6$ field is defined at the scale of the Voronoï cells computed from the center of mass of the drops.
        The grain boundaries are defined by the cells where $\vert\psi_6\vert<0.6$. 
        {\bf B.} Variations of the pair-correlation function $g(\mathbf r)$ along the local principal axis of the crystals.
        The hydrodynamic flows hardly alter the inner structure of the crystallites.
        {\bf C.} Variations of the local magnitude of the $\psi_6$ order parameter in our experiments and simulations. 
        Upon decreasing the magnitude of the repulsive magnetic interactions, the system transitions from an ordered hexagonal state to a disordered liquid state. The white lines indicate the position of the grain boundaries.
        }
        \label{fig:fig2}
    \end{figure*}

    \subsection*{Main findings: Living grain boundaries and  hydrodynamic melting} 

        Before delving into a more thorough discussion, we first offer a bird’s-eye view of our main findings.
        We invite the reader to watch SI videos 1 and 2 first to have a clear visual impression of the polycrystals' structure and dynamics. 
        We summarize them in the series of snapshots shown in Figure~\ref{fig:fig2}.
        For the strongest value of the magnetic field ($B=10\,\rm mT$), the fluid flow leaves the crystal order unchanged, Figure~\ref{fig:fig2}A.
        However, despite the uniform drive, the polycrystals  are not merely advected at constant speed.
        They feature a lively inner dynamics 
        that continuously remold the grain-boundary network, the orientation of the crystal patches, and the distribution of their point defects.
        In the next sections we explain why and how the action of hydrodynamic interactions on dislocations powers and sustains their gliding dynamics.   
        
        Supplementary Movie 2 and  Figures~\ref{fig:fig2}B illustrate our second main finding.
        They show four steady states corresponding to decreasing values of the magnetic field. 
        Although temperature is irrelevant in our system of non-Brownian droplets, we observe that crystalline order is progressively lost as $B$ decreases. 
        The grain size shrinks, while the level of orientational order and the lattice spacing within the crystallites remain unchanged, until they vanish.
        In the next sections, we show that non-reciprocal forces are responsible for a two-way coupling between the crystal structure and dynamics, and explain how they lead to the proliferation of dislocations and grain boundaries.
        \begin{figure*}[ht]
        \centering
        \includegraphics[width=0.9\textwidth]{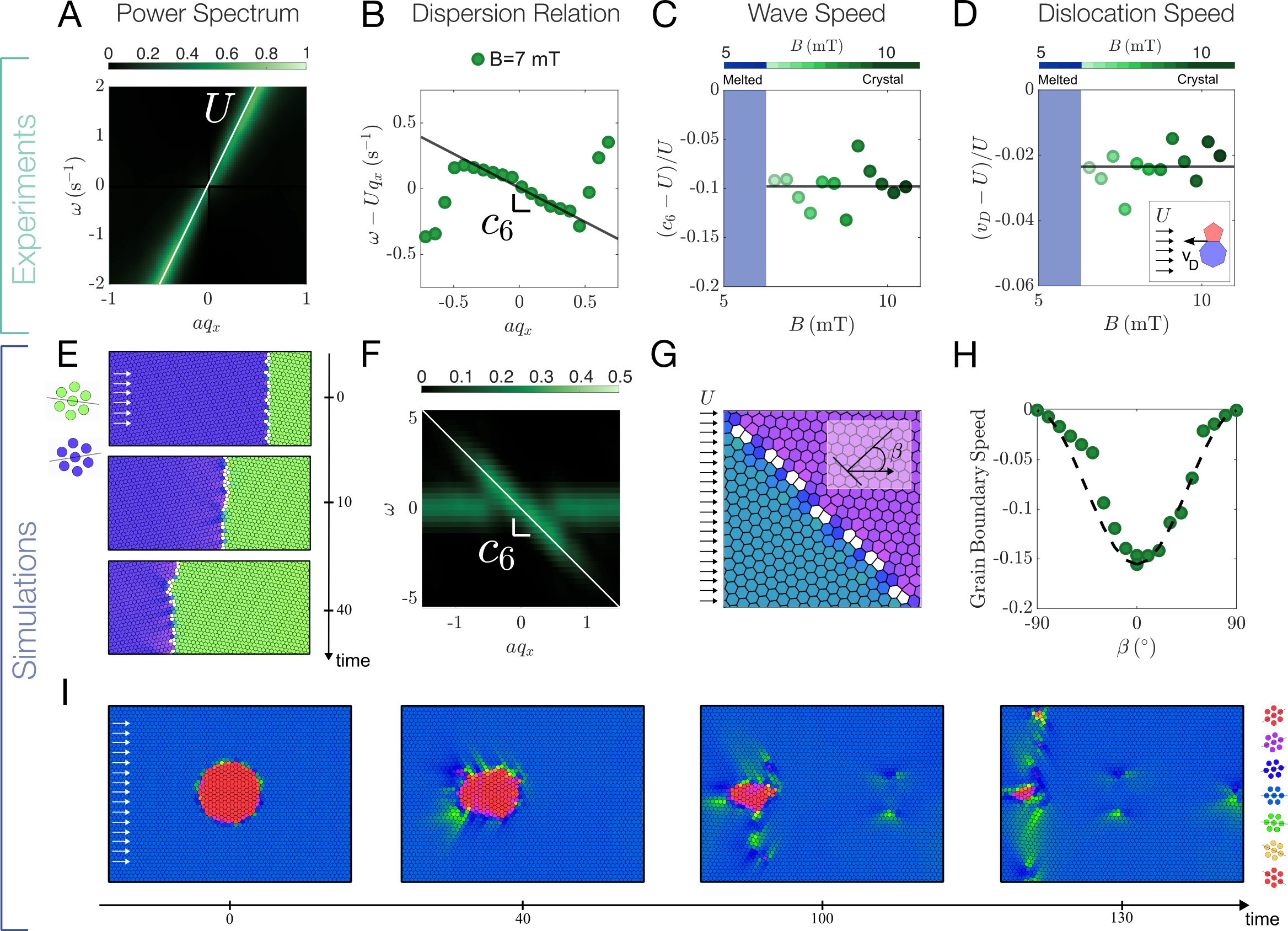}
        \caption{\textbf{Orientational waves and dislocation glide.}
        {\bf A.} Power spectrum of the $\psi_6$ field $\left \langle|\psi_6(q_x,\omega)|\right|^2\rangle$, where $q_x$ represent the wave vectors in the flow direction and $\omega$ the time frequency. 
        The power spectrum is peaked on a line in the $(q_x,\omega)$ plane.
        {\bf B.} This line is however not straight and defines the dispersion relation of the orientational waves. 
        The wave speed corresponds to its slope at the origin and we measure it in the frame moving with the average droplet speed $U=150\pm1\,\rm \mu m/s$.
        Experiments performed at $B=7\,\rm{mT}$
        {\bf C.}
        $c_6$ is negative for all $B$ values: in the comoving frame, orientation waves propagate upstream. 
        {\bf D.} We define the speed of the 
        disclinations $v_{\rm D}$ as the location of the peak of their velocity distribution along the $x$ axis. 
        $v_{\rm D}$ is systematically smaller than $U$ in the polycrystal phase.
        {\bf E.}  Simulation of the dynamics of a straight grain boundary ($\mathcal B=0.2$). The interface between the two crystals is stable but propagates upstream.
        {\bf F.} $\psi_6$ power spectrum corresponding to the numerical simulations of {\bf E}.  Hexatic waves propagate upstream.
        {\bf G.} and {\bf H.} Speed of the straight grain boundary plotted against its orientation with the mean flow (Simulations). The dashed line corresponds to our theoretical prediction ($-\cos^\beta$ function).
        {\bf I.} Simulation of a circular grain. The grain shape is not stable, it undergoes self-sustained deformations powered by the non-reciprocal forces ($\mathcal B=0.1$). 
        }
        \label{fig:fig3}
    \end{figure*}
    
\subsection*{Numerical simulations}
    Before quantifying and explaining our experimental findings,
    we first show that our observations do not depend on the specifics of our experiments, and are generic to the competition between stabilizing repulsive interactions and dipolar non-reciprocal forces.
    To do so, 
    we introduce a minimal model that correctly captures all our  experimental findings.
    We consider the overdamped dynamics of a collection  of point particles labeled by their 2D position $\RR_i(t)$ in a periodic domain. 
    In the frame moving at the average particle speed, their equations of motion takes the simple form:
    \begin{equation}
        \zeta \partial_t {\mathbf R_i}(t)=\sum_j\FF_\mathrm{core}({\mathbf R_{ij}})+\FF_\mathrm{mag}({\mathbf R_{ij}})+\FF_{\rm hydro}({\mathbf R_{ij}}),
        \label{eq:equationmouvement}
    \end{equation}
    where $\zeta$ is a friction coefficient and ${\mathbf R_{ij}}=\mathbf R_i-\mathbf R_j$. In all that follows, units are chosen so that $\zeta=1$.
    The contact interactions between the droplets are accounted for by a short range repulsive force $\FF_{\rm core}$ (WCA potential) that defines the particles' radius $a$.
    We model the magnetic interactions by the repulsion between point dipoles oriented in the direction transverse to the $xy$-plane: $\FF_\mathrm{mag}(\rr) = -{\mathcal B}\nabla \|\rr\|^{-3}$, where $\mathcal B$ quantifies the repulsion strength. 
    Finally, guided by previous models of interacting microfluidic droplets~\cite{Beatus_2006,Desreumaux_2013}, we ignore all near-field hydrodynamic contributions and describe the hydrodynamic interactions by their far-field  component (Figure~\ref{fig:fig1}D):
    $\FF_\mathrm{hydro}(\rr) = \nabla \frac{x}{2\pi r^2}
    $ that has an in-plane dipolar symmetry. 
    Although $\FF_\mathrm{hydro}(\rr)$ can be written as a gradient of a scalar quantity, we stress that it does not obey Newton's third law as $\FF_\mathrm{hydro}(\RR_{ij})=\FF_\mathrm{hydro}(\RR_{ji})$. $\FF_\mathrm{hydro}$ is non-reciprocal. 
    We solve \eqref{eq:equationmouvement} numerically using a modified Ewald algorithm detailed in SI. 
    In all our simulations we set the value of $a$ so that the equivalent packing fraction 
    matches our experimental value. 
    We are thus left with a single parameter $\mathcal B$ that plays the same role as the magnitude of the $B$ field in our experiments.
    It quantifies the relative magnitude of  the reciprocal and non-reciprocal interactions.

    The repulsive interactions organize the particles into a Wigner polycrystal.
    However, in agreement with our experiments, 
    even when the repulsive forces overcome the non-reciprocal interactions, the polycrystals continuously reorganize the shape of their grain boundaries, and their topological defect distributions,  Figure~\ref{fig:fig2}A. Further decreasing $\mathcal B$, the grains' size shrinks until the point where crystalline order is fully destroyed at all scales, Figure~\ref{fig:fig2}C.
    
    The consistency between our experimental and numerical observations unambiguously confirms that the inner dynamics and melting of our driven crystals originate from the competition between potential and non-reciprocal interactions.

    \section*{Results}
    \subsection*{Non-reciprocal interactions propel and remold grain boundaries}
        Our first goal is to quantify and explain the self-sustained dynamics of the grain boundary network in the ordered/polycrystal phase (Figure~\ref{fig:fig2}A and Supplementary Video 1).
        By definition the grain boundaries separate regions of incompatible crystal orientations. 
        We can therefore quantify their dynamics by  measuring the fluctuation spectrum of the orientational (hexatic) order parameter $\psi_6$ , see also SI.
        Figure~\ref{fig:fig3}A shows the spectrum corresponding to wave vectors $\mathbf q=q \hat{\mathbf x}$ pointing along the flow direction. It is peaked on a non-horizontal curve, which reveals that the fluatuations of the hexatic order parameter are not merely overdamped but propagate.
        The $\psi_6$ waves, however, do not merely reflect the homogeneous advection of the polycrystal. 
        We  plot their dispersion relation in Figure~\ref{fig:fig3}B and find that the speed of the orientational-wave $c_{6}$ (the slope at the origin of the dispersion relation) is always negative, Figure~\ref{fig:fig3}C. 
        In other words,
        the fluctuations of the crystals' orientation, and therefore of the grain boundaries, propagate  against the fluid flow. 
        This finding is further confirmed by  tracking the trajectories of the disclinations defined as the Voronoï cells having a number of edges smaller or larger than six.
        The distributions of the disclination velocity are broad (see SI). But, in the longitudinal direction, the speed peaks at a value which is again systematically smaller than the average drop speed  $U$, Figure~\ref{fig:fig3}D.
        The elementary topological defects propagate upstream, however they do not merely translate but also continuously reorganize the structure of the grains.
        To elucidate this complex dynamics we now take advantage of our numerical simulations.
        Unlike in experiments, we can prepare two pristine Wigner crystals separated by two straight grain boundaries orthogonal to the flow. 
        We can then let them evolve under the action of the non-reciprocal forces.
        In Figures~\ref{fig:fig3}E, ~\ref{fig:fig3}F and Supplementary video 3, we observe that the grain boundaries and the associated $\psi_6$ fluctuations propel steadily in the upstream direction.
        We then repeat this measurement for grain boundaries forming a finite angle $\beta$ with the flow as illustrated in Figure~\ref{fig:fig3}G. 
        We find that their shape remains unchanged but that their speed vary with $\beta$, see Fig.~\ref{fig:fig3}H. 
        Consequently curved grain boundaries must deform %as they are propelled by non-reciprocal forces, 
        as confirmed by the simulations of a circular grain showed in Figure~\ref{fig:fig3}I and Supplementary video 4. 
        
        Our recent theory exposed in Refs.~\cite{Poncet2022}, provides an explanation for the motility of the grain boundaries.
        We recall and generalize it in SI. 
        In short, when a dislocation of Burgers vector $\mathbf b$  forms in a crystal, it causes elastic deformations at all scales. 
        The resulting strain induces non-reciprocal interactions between the particles that build  an effective  Peach-K\"ohler force $\mathbf F^{\rm PK}$ acting on the dislocation~\cite{Oswald,Poncet2022}. 
        $\mathbf F^{\rm PK}$ decomposes into a glide $F^{\rm PK}_{\rm glide}=\mathbf F^{\rm PK}\cdot\hat{\mathbf b}$ and a climb component $F^{\rm PK}_{\rm climb}=\mathbf F^{\rm PK}\cdot\hat{\mathbf b}^{\perp}$.
        When the non-reciprocal forces originate from dipolar hydrodynamic interactions ($\FF_{\rm hydro}$), the total Peach-K\"ohler force points along the flow direction and the glide component reduces to
        \begin{equation}
        {\mathbf F^{\rm PK}_{\rm glide}\propto -(\mathbf b\cdot\hat{\mathbf x})\bb}.
        \label{PKglide}
        \end{equation}

        This relation captures all the ingredients needed to explain our first set of experimental and numerical observations: (i)
        $\mathbf F^{\rm PK}_{\rm glide}$ points in the direction opposite to the driving flow,  it therefore powers the upstream motion of any dislocation whose burgers vector does not make a $\pi/2$ angle with the horizontal direction. 
        (ii) The magnitude of the Peach-K\"ohler force  depends on the orientation of the dislocations.  
        In agreement with our experimental and numerical results reported in Figure~\ref{fig:fig3}, ignoring the effect of elastic interactions between the dislocations,
        \eqref{PKglide}  indicates that straight grain boundaries should propagate steadily against the macroscopic drive, whereas curved grains should deform under the action of the hydrodynamic interactions. 
         \begin{figure*}[ht]
            \centering
            \includegraphics[width=0.9\textwidth]{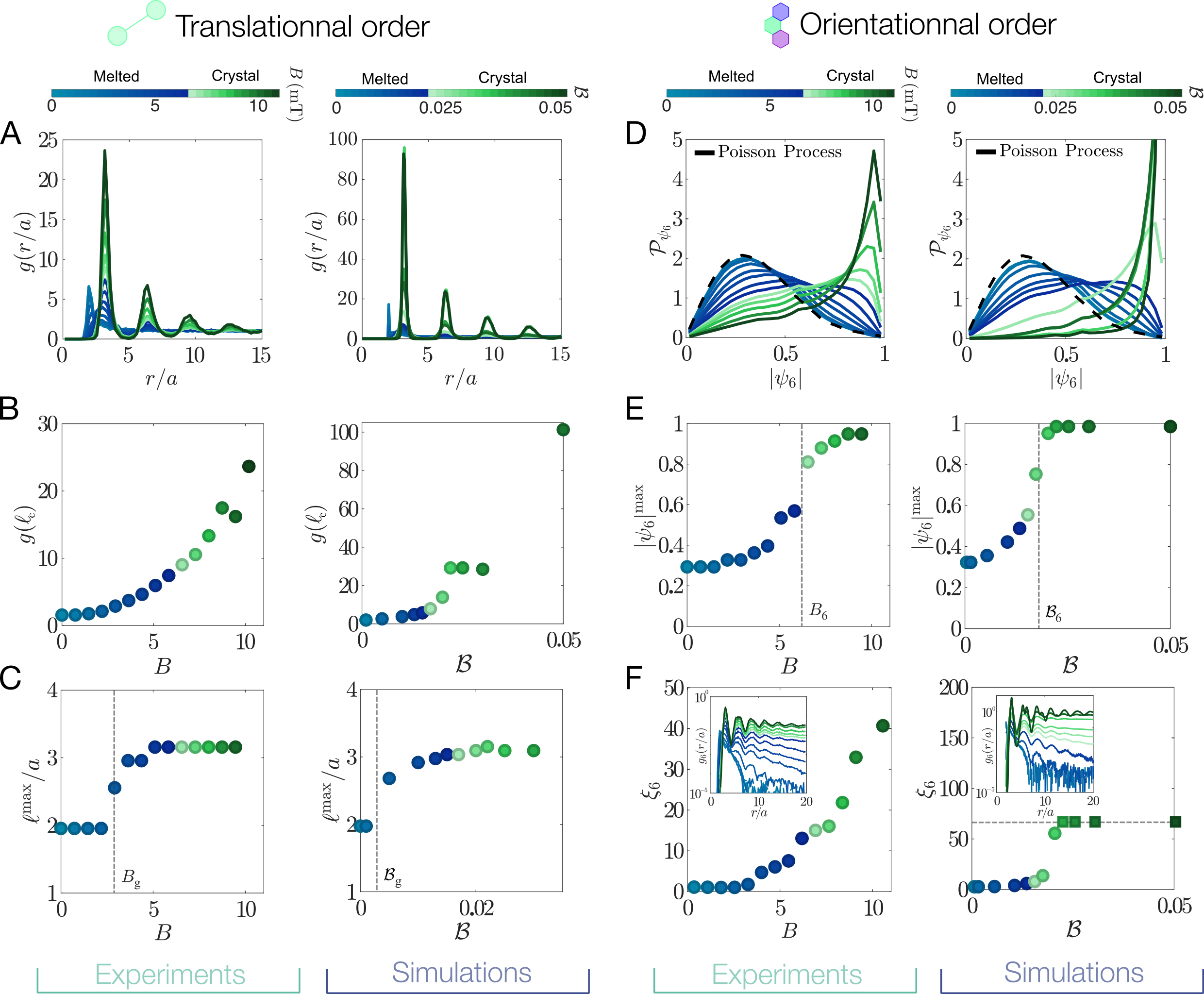}
            \caption{\textbf{Hydrodynamic melting.} 
            {\bf A.} Pair correlation functions $g(r/a)$ measured along the direction defined by the local value of $\theta_6$. $g(r/a)$ is plotted for different values of the applied magnetic field. 
            The structure evolve from a crystal to a liquid. Experiments and Simulations.
           {\bf B.} Variations of the amplitude of the first crystal peak $g(\ell_{\rm c})$ with $B$. The position $\ell_{\rm c}$ of the first crystal peak is defined in panel {\bf A}. 
           Experiments and Simulations.
           {\bf C.} The position of the first peak of $g(r/a)$ ($\ell^{\rm max}$) does not vary with over a range of $B$ values. The structure of the Wigner crystals does not change in the bulk as $B$ decreases. Experiments and Simulations.
           {\bf D.} Distribution of the magnitude of the local and instantaneous hexatic field $\psi_{6}(\mathbf r,t)$. Dashed line same distribution computed for a random set of points (2D Poissonian process). Experiments and Simulations.
           {\bf E.} Evolution of the position of the maximum of the $\psi_6$ distribution with the magnitude of the repulsive interactions. Experiments and Simulations. The position of the $\psi_6$ maximum clearly signals the melting point of the polycrystal structure.
            {\bf F.} The correlation lengths $\xi_6$  associated with the spatial correlations of $\psi_6$ (Inset) continuously decreases as the strength of the repulsive force decreases and the polycrystal melts. Experiments and Simulations. 
            In our simulations $\xi_6$ plateaus at high $\mathcal B$ values as it reaches the system size.}
            \label{fig:fig4}
        \end{figure*}

    \subsection*{Nature of the hydrodynamic melting of Wigner polycrystals}
       We have demonstrated that non-reciprocal interactions sustain the continuous remolding of domain walls in hydrodynamically driven crystals.
        We now ask how non-reciprocal interactions disrupt crystalline order as elastic interactions weaken (Figure~\ref{fig:fig2}C).
        To answer this question, we first  characterize the evolution of translational and orientational order parameters in  steady state as we gradually decrease the magnitude of the magnetic forces.  
        We then explain our findings, and explain  how the hydrodynamic interactions  promote dislocation splitting and govern the melting process.
        
        To characterize the structural evolution of our emulsions, we plot  the pair correlation functions of the droplets for different values of the $B$ field (experiments and simulations),  Figure~\ref{fig:fig4}A. 
        As the elastic forces weaken, the amplitude of the crystal peaks decreases continuously, Figure~\ref{fig:fig4}B.
        However, at high $B$, the locations of the  peaks do not vary, and no new structural feature emerges: the local symmetry and period of the  crystals  remain the same, Figure~\ref{fig:fig4}C.
        This observation is at odds with the clumping-induced-melting scenario reported in various hydrodynamic crystals, from the early experiments of Crowley~\cite{Crowley71}, to the numerical simulations of Ref.~\cite{Saeed2023} (see also~\cite{Lahiri_1997,Ramaswamy2001}).
        Further decreasing $B$, the structure of the emulsion then changes sharply, Figures~\ref{fig:fig4}C.
        When $B<B_{g}=3.25\,\rm{mT}$, no crystallite survives, the crystal peaks vanish, and we are left with a fully disordered liquid whose structure hardly depends on $B$, Figure~\ref{fig:fig4}A.  
        
        The evolution of the orientational order parameter provides another insight into the melting process.
        In Figure~\ref{fig:fig4}D, we plot the distributions of magnitude of the local $\psi_6$ field. 
        When the elastic interactions are strong (high $B$), the distributions peak at a value close to one. 
        The location of the $\psi_6$ peak hardly shifts towards smaller values until $B=B_6=6.4\,\rm mT$  (experiments), Figure~\ref{fig:fig4}E. 
        It then discontinuously drops to reach a value close to the maximum of the $\psi_6$ distribution computed for a Poisson point process.
        These variations provide a clear definition of the crystalline clusters: they correspond to regions where the value of $\psi_6$ exceeds $0.6$. 
        In Figure~\ref{fig:fig4}F, we plot the correlation length of the orientational order parameter $\xi_6$ against $B$. 
        We find that, for strong magnetic repulsion, $\xi_6$ continuously decreases while the magnitude of the orientational order parameter  remains nearly constant (Figures~\ref{fig:fig4}B and ~\ref{fig:fig4}F).
        We therefore conclude that the loss of orientational order does not happen within the crystal patches, but reflects their  shrinking as $B$ decreases. 

        As the hydrodynamic interactions prevail over the elastic forces, a sea of liquid phase grows around nearly pristine crystallites whose  size shrinks in response to   non-reciprocal interactions. 
    
\begin{figure*}[hbtp]
            \centering
            \includegraphics[width=\textwidth]{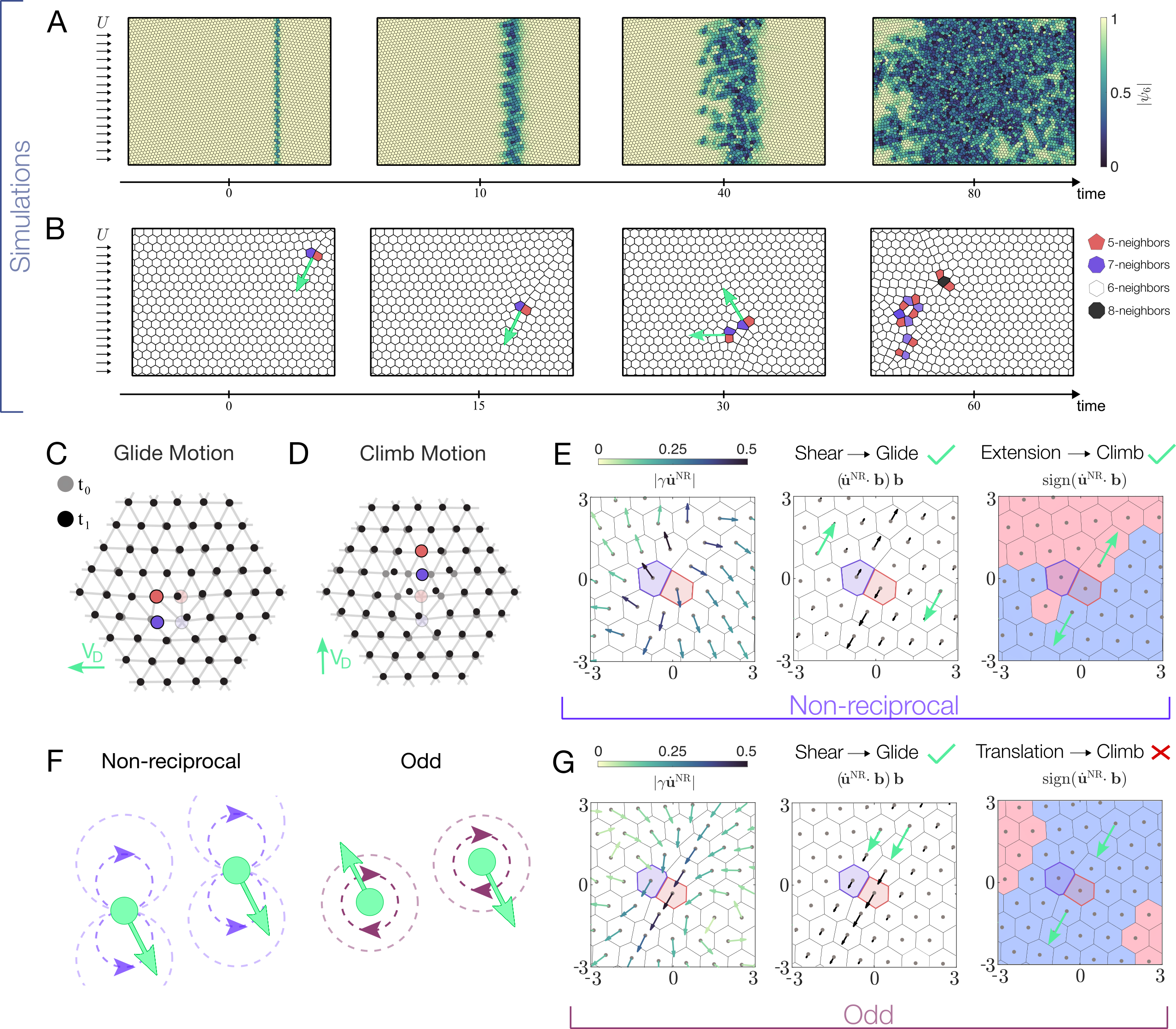}
            \caption{\textbf{Nonreciprocal forces split dislocations.} 
            {\bf A.} Structural evolution of two crystals separated by a straight grain boundary upon a rapid quench. 
            We initialize the system by relaxing the elastic energy ($\mathcal B=0$), we then turn on  the nonreciprocal interactions ($\mathcal{B}=5\times 10^{-3}$) and let the polycrystal evolve in time.
            The dislocations forming the interfaces splits, propel, and react while the bulk of the crystals remains unperturbed. The dislocations then proliferate across the whole sample leading to a dynamical steady states where small and lively grains are surrounded by a thick liquid phase. See also Supplementary Video 5.
            {\bf B.} Simulation of the dynamics of a single dislocation under the action of repulsive  and nonreciprocal forces ($\mathcal{B}=5\times 10^{-3}$).  The dislocation first glides, it then splits into a pair instead of climbing. The dislocation pair  propels, interacts, splits again leading to their proliferation and the  destruction the initial crystal order. See also Supplementary Video 6.
            {\bf C. and D.} Schematics of glide and climb motion
            {\bf E.} (Left panel) Force field induced by pairwise additive nonreciprocal forces having the same dipolar symmetry as in our experiments. 
            The force field is computed in the vicinity of an isolated dislocation.
            (Middle) The arrows represent the projections of $\mathbf F^{\rm NR}=\gamma\dot{\mathbf u}^{\rm NR}$ on the Burgers vector $\mathbf b$. 
            A simple shear powers the glide of the dislocation. 
            (Right)
            The colors indicate the sign of the projection of the non-reciprocal forces on the $\mathbf b$ vector.
            The sign of the force changes across the dislocation thereby promoting a climbing dynamics. 
            {\bf F.} Illustration of pairwise nonreciprocal and odd forces. Nonreciprocal forces do not feature the action reaction symmetry, but result in a vanishing torque. Conversely odd forces satisfy Newton's action-reaction symmetry, but their transverse direction induces a net torque on a pair of particles. 
            {\bf G.} Same plots as in {\bf E} for an odd crystal. The collection of particles interact via the same conservative repulsion and are driven out of equilibrium by  transverse interactions modeled by pairwise additive forces $\mathbf F^{\rm odd}(\mathbf r)=e^{-r}{\mathbf r}^{\perp}$. 
            (left) The odd force field forms two visible vortex around an isolated dislocation. 
            (Middle) The projection of the odd-force field on $\mathbf b$ reveals a net simple shear (i.e a non vanishing vorticity) acting on the dislocation, it causes its gliding dynamics. 
            (right) Unlike in the transverse direction the off-force field hardly varies along $\mathbf b$. Odd forces cannot power any climb dynamics.   
            }
            \label{fig:fig5}
        \end{figure*}

        To further confirm the prominent role played by grain boundaries in our melting process, we take advantage of our numerical simulations. 
        We start with two crystals separated by two straight interfaces, and quench the system right below the ``melting point'', $\mathcal B=17\times10^{-3}$ in our simulations. 
        Figure ~\ref{fig:fig5}A clearly reveals that topological defects hardly ever nucleate in the bulk but proliferate from the grain boundaries. 
        Ultimately, in agreement with our experimental observations, the defect proliferation leads to a macroscopic reshaping of the polycrystal geometry and leads to the formation of small ordered clusters that coexist with macroscopic disordered regions.
        
        To elucidate this melting process, we consider the even simpler case of an isolated dislocation, see Fig.~\ref{fig:fig5}B. 
        When $\mathcal B$ is set right below the melting point $\mathcal B_6$, we find that at early times, the dislocation fissions to form a pair of  gliding defects.
        We can understand this fission process as follows. 
        The sum of the hydrodynamic interactions acting around a dislocation results in an effective Peach-K\"ohler force that includes a climb component:
        \begin{equation}
            {\mathbf F^{\rm PK}_{\rm climb}\propto -(\mathbf b^\perp\cdot\hat{\mathbf x})\bb^\perp}
        \label{eq:Fclimb}
        \end{equation}
        where $\mathbf b$ is the Burgers vector.
        However,   a dislocation cannot propel along the climb direction due the extensive energy cost associated with their displacement.
        In agreement with the optical tweezer experiments reported in Ref.~\cite{Irvine_2013}, we find that  above a critical local strain,  an isolated dislocation reacts by fissioning into  a defect pair, keeping the overall topological charge constant, see Fig.~\ref{fig:fig5}B.
        At later times, the propulsion of the dislocation pair, their interactions, and subsequent fission reactions result in a complex chain reaction that gradually destroy the crystal ordering.
        
        We stress that the defect splitting process rules the melting dynamics, as the bulk of our hydrodynamic crystal is stable. 
        In a perfect periodic lattice the sum of all hydrodynamic interactions cancel out exactly, and governs a dynamics that is is linearly stable to structural fluctuations, see~\cite{Desreumaux_2012,Saeed2023} and SI.

       In sum, the melting process that we have uncovered is typical of driven non-equilibrium matter whose inner structure and dynamics cannot be disentangled.
        As  dislocations form, the  changes in the crystal structure induces  hydrodynamic forces that alter the particle dynamics and fluid flows. 
        The flows  cause the fission of the dislocations. These structural changes then feedback again on the particle dynamics thereby leading to a cascade of fission and annihilation events that dynamically shape the structure of the polycrystals.
        
        This melting process is markedly different from the entropic proliferation of topological defects in the bulk of thermal 2D crystals~\cite{kardar2007fields}.
        It is however reminiscent of the nucleation of dynamical crystals in collections of dense self-propelled hard disks~\cite{Briand_2016}, and the suppression of order from living grain boundaries is strikingly similar to the melting of odd colloidal crystals reported in~\cite{Bililign2022}. 
        This resemblance is not anecdotal and hints towards seemingly similar mechanisms which we discuss in the next  section. 
        
\section*{Discussion}
\subsection*{Beyond hydrodynamic interactions: Dislocation motion and proliferation in non-reciprocal crystals}

Returning to the opening question of this article, we now provide general theoretical arguments to demonstrate that our findings are not specific to driven emulsions:  microscopic forces that violate Newton's third law generically promote the glide and splitting  of dislocations.

    To see this,  we first recall the basic kinematics of dislocation dynamics, see  Figures~\ref{fig:fig5}C and~\ref{fig:fig5}D. 
    Consider a collection of particles self-organized into a periodic lattice including an isolated dislocation of Burgers vector $\mathbf b$.
    The glide of a dislocation over one lattice spacing results from minute relative displacements of the particles along the $\mathbf b$ direction, right above and  below the dislocations. 
    In other words, the gliding speed of a dislocation is proportional to the local shear rate across the topological defect. 
    At a continuum level,  a dislocation located at $\mathbf r=0$ glides at a speed 
    \begin{equation}
        v_{\rm glide}\propto
        \mathbf b^{\perp}\cdot\bm \nabla (\mathbf b\cdot\dot{ \mathbf u})|_{\mathbf r=0},
        \label{eq:glide}
    \end{equation}%
    where $\mathbf u(\mathbf r,t)$ is the local-displacement field. 
    Similarly, the climb speed of a dislocation is  set by the local compression rate of the lattice along the axis defined by its Burgers vector, Figure~\ref{fig:fig5}D, and $v_{\rm climb}\propto\mathbf b\cdot\bm \nabla (\mathbf b\cdot\dot{ \mathbf u})|_{\mathbf r=0}$. 

    To quantify the deformation rate of a crystal, we need to prescribe its dynamics. 
   The displacement field of driven crystals  obeys the generic momentum conservation equation: 
    $ D_t(\rho \dot{\mathbf u} )=\bm \nabla \cdot \bm \sigma+\mathbf F(\{\mathbf u(\mathbf r)\},\mathbf r)$,
    where $\rho$ is the mass density, $D_t$ is the convective derivative, and $\bm \sigma$ the stress tensor that originates from conservative interactions at the microscopic scales.
    When the body force acting on the crystal does not depend on the particle arrangement, $\mathbf F$ classically reduces to the standard  sum of a dissipative drag $-\gamma \dot{\mathbf{u}}(\mathbf{r},t)$
    and of a driving force $\mathbf F_0(\mathbf r)$.
   This situation is relevant to  systems as diverse as  bubble rafts mechanically sheared in a quasistatic regime~\cite{bragg1947}, or  Abrikosov lattices driven by an electric field and pinned by quenched disorder~\cite{Vinokur94}.
 
   However, as noted first in~\cite{Lahiri_1997}, the exchange of momentum between the driven particles and their surrounding environment can result in a driving force field that  depends on the particles' configuration. 
   We note it $\mathbf F^{\rm NR}(\lbrace\bm \nabla\mathbf u(\mathbf r,t)\rbrace)$. 
It is relevant when particles interact, for instance, through hydrodynamic or phoretic interactions. 
As such microscopic forces  violate Newton’s third law, they act as sources or sinks of linear momentum~\cite{Poncet2022}.
$\mathbf F^{\rm NR}$ has a simple but counter-intuitive consequence: a homogeneous deformation  results in a net acceleration of the crystal, but a global acceleration does not yield any deformation.
Having in mind soft materials assembled from microscopic constituents, we  focus on the overdamped regime where the drag force dominates inertia and simplify the equations of motion as
    \begin{equation}
        \gamma \dot{u}_i
        =\partial_j \sigma_{ij}+F^{\rm NR}_i(\lbrace \bm\nabla \mathbf u\rbrace ).
        \label{eq:overdamped}
    \end{equation}
    In this  limit, the local velocities are proportional to the  local forces;  the action of the non-reciprocal interactions can always be seen as the result of the advection of the particles  by an effective flow ${\dot{\mathbf u}^{NR}}=\gamma^{-1}\mathbf F^{NR}$. 
    When they arise from hydrodynamic interactions $\dot{\mathbf u}^{NR}$ is directly proportional to the solvent backflows caused by the motion of the particles, and $\mathbf F^{\rm NR}$ can be thought as a strain dependent drag force~\cite{Lahiri_1997}.

    We are now one step away from showing that $\mathbf F^{NR}$ powers the dynamics of dislocations.
    As a matter of fact, in elastic crystals,  dislocations result in  non-vanishing shear, dilation and rotation strains whose  magnitude decay algebraically in space ($\nabla\mathbf u(\mathbf r)\sim 1/r$)~\cite{Landau}. 
    As  no conservation law, or symmetry principle, require $\nabla b_i \dot{u}^{\rm NR}_i$ to vanish, we readilly conclude from \eqref{eq:glide} that microscopic forces violating the action-reaction principle induce and sustain local displacements that promote both the glide and climb motion of dislocations:
    dislocations in non-reciprocal crystals are self-propelled.

    To gain more intuition, in SI, we show how to  compute $\mathbf F^{\rm NR}$, and the associated dislocation speed in two cases: 
    (i) When the nonreciprocal interactions are  short ranged $\mathbf F^{\rm NR}$ then takes the generic form $ F^{\rm NR}_i=A_{ijk}\partial_ku_j(\mathbf r)$ where $A_{ijk}$ depends both on the symmetries of the forces and of the lattices, see also~\cite{Poncet2022}. 
    (ii) When the nonreciprocal interactions are
    the long-range hydrodynamic forces ($\mathbf F_{\rm hydro}$) relevant to our experiments we find a simple non-local relation that reduces to the convolution $\mathbf F^{\rm NR}\sim-\mathbf F_{\rm hydro}*(\bm\nabla\cdot \mathbf u)$.
     We further illustrate our reasoning  in Figure~\ref{fig:fig5}E.
    We consider an isolated dislocation in an elastic crystal whose the dynamics is ruled by \eqref{eq:equationmouvement}, and show the corresponding non-reciprocal flow $\dot{\mathbf u}^{\rm NR}$, along with its projection on the Burgers' vector direction, and its sign.
    In agreement with  analytical calculations reported in SI, both    $\mathbf b\cdot\bm\nabla(\mathbf{b}\cdot\dot{\mathbf u}^{\rm NR})$ and $\mathbf b^\perp\cdot\bm\nabla(\mathbf{b}\cdot\dot{\mathbf u}^{\rm NR})$ take finite values in the vicinity of the dislocation core, thereby promoting both its glide and climb motion. 

    As discussed in the previous section, we stress  that the glide and climb dynamics have a very different status in elastic crystals. 
    Unlike gliding, climbing is practically never observed due to its extensive elastic energy cost.
    Instead of climbing, it is energetically favorable to fission the dislocation into two defects that both glide, thereby resulting in an effective climb of the net topological charge of the dislocation pair, see Figure~\ref{fig:fig5}B and~\cite{Irvine_2013}. 
   This proliferation process  ultimately promotes the melting of elastic crystals perturbed by strong non-reciprocal forces, regardless of their microscopic origin.
    \subsection*{Dislocation motion and proliferation in odd crystals}
    We now close our discussion by commenting on the  similarities and fundamental differences with another form of non-equilibrium matter: odd crystals assembled from active spinners. 
   Prominent examples include shaken chiral grains, magnetically-driven colloids and oocyte eggs~\cite{Scheibner_2020,Bililign2022,Tan2022,fruchart2023odd}.
    In odd crystals, the elementary units  interact through forces that satisfy Newton's action-reaction symmetry.
    But they act in the direction transverse to the separation vectors, Figure~\ref{fig:fig5}F:
    a pair of particle coupled by odd/transverse forces  experience a net torque. 
    Experiments and simulations revealed that the grain boundaries of odd polycrystals feature a lively dynamics strikingly similar to that observed in our experiments~\cite{Bililign2022,caporusso2024phase}. 
    This dynamics can be understood within the framework we have introduced above. 
    In short, spinners are driven by local torque sources. 
    As soon as the particles are coupled by finite frictional interactions,
    this local injection of angular momentum at the microscopic level promotes the emergence of a non-vanishing vorticity  $\Omega=\frac 1 2\epsilon_{ij}\partial_i \dot{\mathbf u}_j$
    , see e.g.~\cite{Tsai_2005,Bonthuis_2009,Soni2019}.
    This net  vorticity is not obvious when inspecting the full flow field of Figure~\ref{fig:fig5}G, but it becomes clear when looking at its projection in the direction defined by the Burgers vector.
    We can readily see that odd interactions result in a  net velocity gradient across an isolated dislocation. 
    Equation \eqref{eq:glide} then reveals that this kinematics drives the gliding motion of dislocations.
    Crucially, however, unlike non-reciprocal interactions, odd forces do not promote the climb dynamics of isolated dislocations.
    Figure~\ref{fig:fig5}G indeed shows that the odd forces do not result in any flow gradient in the Burgers vector direction, \eqref{eq:glide}. This implies that the dislocation cannot climb.
    The proliferation of dislocations and the subsequent fragmentation of odd crystals is therefore fundamentally different from the non-reciprocal melting dynamics which we characterize and explain in Figures~\ref{fig:fig3} and~\ref{fig:fig4}. 
    Transverse interactions cannot compete with elastic forces to fission isolated topological defects. 
    The microscopic mechanisms leading to the fragmentation of odd crystals must involve  interactions between motile dislocations which remain to be elucidated.
    
\section*{Conclusion and outlook}
    Combining experiments and numerical simulations, we have shown that  interactions  that do not obey Newton's action-reaction symmetry propel and fission dislocations hosted in soft crystals. 
    At macroscopic scales, these microscopic dynamics translate into the self-sustained remolding of grain-boundary networks and  to the  destruction of crystal order.
    Our theoretical analysis based on conservation laws and symmetry principles reveals that these phenomena are universal and should be observed in any non-equilibrium crystals challenged by non-reciprocal forces. 
    Our work focuses on hydrodynamically driven crystals, but we provide theoretical evidence for the robustness of our findings. These insights should apply to a much broader class of nonequilibrium systems, where topological defects and non-reciprocal interactions conspire to shape the structures and dynamics of interacting units, ranging from chemically active colloids to living cells and coupled oscillators.

\begin{acknowledgments}
\textbf{Acknowledgments} This work was partly supported by the European Research Council (ERC) under the European Union’s Horizon 2020 research and innovation program (grant agreement No. [101019141]) (DB).
We thanks Ephraim Bililign, Yehuda Ganan, Michel Fruchart, and Colin Scheibner for illuminating discussions.
\end{acknowledgments}

\bibliography{./biblio.bib}

%%%%%%%%%%%SUPPLEMENTARY INFORMATIONS%%%%%%%%%%%%%%%%

\end{document}

\title{Supplemental Material: \\
Melting of non reciprocal solids: how dislocations propel and fission in flowing crystals}

\author{St\'ephane Guillet}
\author{Alexis Poncet}
\affiliation{ENS de Lyon, CNRS, LPENSL, UMR5672, 69342, Lyon cedex 07, France}
\author{Marine Le Blay}
\affiliation{Pôle d’Etudes et Recherche de Lacq, Total SA, BP 47-64170 Lacq, France}
\affiliation{ENS de Lyon, CNRS, LPENSL, UMR5672, 69342, Lyon cedex 07, France}
\author{William. T. M Irvine}
\author{Vincenzo Vitelli}
\affiliation{James Franck Institute, University of Chicago, Chicago, IL 60637}
\affiliation{Department of Physics, University of Chicago, Chicago, IL 60637}
\author{Denis Bartolo}
\affiliation{ENS de Lyon, CNRS, LPENSL, UMR5672, 69342, Lyon cedex 07, France}

\maketitle

%%%%%%%%%% Prefix a "S" to all equations, figures, tables and reset the counter %%%%%%%%%%
\setcounter{equation}{0}
\setcounter{figure}{0}
\setcounter{table}{0}
\setcounter{page}{1}
%\makeatletter
\renewcommand{\thefigure}{S\arabic{figure}}
\renewcommand{\theequation}{$\mathcal{S}$\arabic{equation}}
\renewcommand{\bibnumfmt}[1]{[S#1]}
\renewcommand{\citenumfont}[1]{S#1}
\renewcommand\thesubsection{\arabic{subsection}}
\renewcommand\thesubsubsection{\arabic{subsection}.\arabic{subsubsection}  }

\section{Experimental methods}
    We first outline the methods we use to make our microfluidc devices. Next, we describe the methods used to make magnetic Winger crystals out of microfludic droplets, and we provide a detailed explanation of the data acquisition and analysis processes.
    \subsection{Micro-fabrication}
        \subsubsection{Microfluidic stickers}
            The fabrication of the microfluidic chip  follows the protocol described in \cite{Bartolo_stickers_2008}. In short, we use conventional UV lithography techniques to make a master mold of our microchannels (SU8 photoresists). 
            We make a PDMS replica of this geometry (Polydimethyl Siloxane, Silguard) and use the resulting soft stamp to imprint another photosensitive resin NOA$81$ (Norland Optical Adhesive) to make a micropatterned sticker.
            We then close the channels by sticking our NOA81 channels on a glass slide of thickness $1\rm{\,mm}$. 
            Before starting a series of experiments, the cell is exposed to intense UV radiation in an ozone atmosphere. This step eliminates organic compounds and makes the channel walls highly hydrophilic~\cite{Levache2012}. The stickers are connected to three injection tubes (ETFE Tefzel tubes of inner diameter $1/16"$) using homemade connectors sealed with epoxy resin at the inlet and outlet of the channel.
        
        \subsubsection{Fabrication of the PDMS Stamp}
            Creating a microfluidic sticker involves illuminating a photosensitive resin through a PDMS stamp. 
            The first step in making the stamp involves creating a master mold through photolithography, following the protocol in \cite{Kayaku_2020}. 
            This model is made by depositing negative SU-8 2050 resin (Microchem) on a silicon wafer. Before this, the wafer is heated to 100°C to remove any residual moisture. Then, the substrate and the resin are spin-coated for 10 seconds at 500 rpm and then for 30 seconds at 4000 rpm. They are then baked at 65°C for 3 minutes and then at 95°C for 9 minutes (soft bake). The mask defining the channel geometry is placed between the wafer and a UV-transparent quartz plate, which is then exposed to UV radiations (365 nm, $30 \rm{\,mW/cm^2}$) for 25 seconds. 
            Then, the resin is cured for 2 minutes at 65°C and then for 7 minutes at 95°C.  The wafer is cleaned to remove any uncured resin by immersing it in a developer bath for 10 minutes. 
            We finalize the fabrication of the master mold by rincing it with isopropanol and  heating it for 3 minutes at 150°C (hard bake), which eliminates micro-cracks smaller than $5 \rm{\,\mu m} $in size. Finally, we pour 30 grams of PDMS Sylgard 184 from Dow Corning (curing agent 10\% mass) the wafer and degas it under vacuum. The mixture is allowed to cure for 2 hours at 72°C. 
            We peal off the PDMS stamp which we use to imprint our $e=60\,\rm{\mu m}$ deep channels in NOA81 resin.

    \subsection{Making a soft Wigner crystal}
\subsubsection{Droplet production}
        The microfluidic cell  includes  a T-junction at the inlet for drop generation and advection. 
        We inject the continuous and the dispersed phases in the two  inlet channels  to form a monodisperse emulsion one droplet at a time at the T-junction. 
        The emulsion is made of Hexadecane droplets dispersed in a continuous phase of ferrofluid (Ferrotec MSG $W11+0.3\%w.t.$ SDS). The saturation magnetization of MSG W11 is $M_s=18.5\,\rm{mT}$. We filter the ferrofluid beforehand using  $0.22\,\rm{\mu m}$ syringe filters without altering its magnetic properties. 
       The droplets are then advected though a tree network of channels to homogeneously fill the main Hele-Shaw cell in which we conduct our observations. 
       To make droplets of controlled diameter, Figure~\ref{fig:figS1}A), we impose the flow rates of the two  fluids with a Nemesys Cetoni precision syringe pumps, see Fig. 
        We sets the area fraction of the droplets to $\phi^1=0.61$ and $\phi^2=0.37$ by imposing the flow rates ($Q^1_{\rm ferro}=\SI{6}{\micro L/\min}$, $Q^1_{\rm Hexadecane}=\SI{1.5}{\micro L/\min}$) and ($Q^2_{\rm ferro}=\SI{25}{\micro L/\min}$ , $Q^2_{\rm Hexadecane}=\SI{2}{\micro L/\min}$) respectively.

\subsubsection{Crystalization}
        
\begin{figure}[t]
            \includegraphics[width=\textwidth]{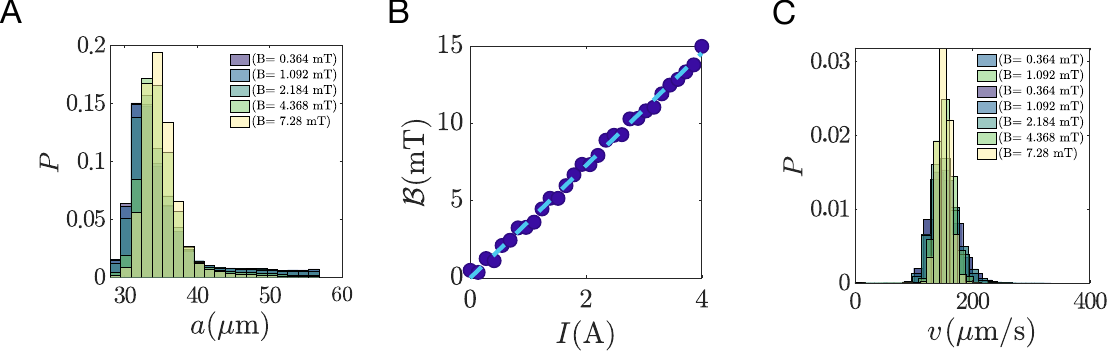}
            \caption{\textbf{Controlling the production of a magnetic soft crystal}
            {\bf A.} The radius of droplets does not vary with $B$.
            {\bf B.}Variations of the $\bf B$ field strength with the intensity of the current in the coil. 
            {\bf C.} Velocity distributions of the droplets for different applied magnetic field. No crucial difference is observed in the velocity distribution of the droplets. 
            }
            \label{fig:figS1}
        \end{figure}

        To organize the droplets into a polycrystal, we use  the protocol introduced in \cite{Sjeltorp_holes_1983}. We place the microfluidic device in a homogeneous transverse magnetic field $\bf B$. 
        We control the strength of the $\mathbf B$ field using an electric generator. 
        The magnetic field inside the coil varies linearly with the applied current $I$, see Figure~\ref{fig:figS1}B). 
         We stress that the mean advection velocity of the crystal is independent of the applied magnetic field (Figure~\ref{fig:figS1}C). 

    \subsection{Data Acquisition \& Analysis}
        \subsubsection{Acquisition}
            We image our droplet crystals using a Nikon $SMZ800$ stereo microscope with a $0.5\times$ achromatic objective. 
            The chips were positioned at the center of a coil and illuminated with a LED light plate. The experiments were recorded using a $12$ Megapixel CMOS camera (Ximea XiC MC124MG-SY). 
            Depending on the injection rates, the duration of the recordings was either set to $\SI{300}{\second}$ ($7$ fps), or $\SI{120}{\second}$ ($10$ fps). 
            We chose these values to allow for droplet trajectory reconstruction. The size of the observation window is $\sim\SI{5}{\milli\meter}\times\SI{2}{\milli\meter}$, representing approximately $\sim10\%$ of the cell, in which we ensure that the density field is relatively homogeneous. Each experiment was conducted with a fixed magnetic field, which was then increased in increments of $\SI{0.364}{\milli T}$ until reaching its maximum value of $\SI{10}{\milli T}$.

           % We employ the same observational setup for studying the dynamic rearrangements of the crystal, with a frame rate set at $2$ fps, covering a typical region of interest of size $16\,\rm{mm}\times10\,\rm{mm}$. Once a stationnary crystalline state is reached, we set the continuous phase flow at $Q=\SI{1}{\micro L/\min}$ for a duration of $20$min, leading to the restructuring of the system.

        \subsubsection{Particle Detection \& Tracking}
           To detect the drop position, we find the local intensity maxima in each image. To detect all the particles, the following pre-processing steps were applied: the image contrast is automatically increased and a Gaussian blur with a size of $2.5$ px applied  to smooth the intensity profile. 
           This provides a list of coordinates for each particle and for each frame (see Figure~\ref{fig:figS2}A). 
           We track the particles and reconstruct their trajectories using the Crocker and Grier method~\cite{Grier_tracking_1996}.
           We show typical trajectories  in Figure~\ref{fig:figS2}B. 
           We compute the Delaunay triangulation diagram and the adjacency matrix based on the particle positions in each frame. 
           This allows to associate each particle to its  nearest neighbors.

        \subsubsection{Translational and orientational order parameter}
           We define two  order parameters. 
            \begin{itemize}
                \item \textbf{Translational order.}  $g({\bf r})$ quantifies the degree of correlation between the particle positions. It is defined as the probability (per unit volume) to find a particle at the position $\mathbf r+\mathbf R$ given a particle at $\mathbf R$ :
                $g(\mathbf{r})=\frac{1}{\langle n\rangle}\left\langle\sum_{j\neq i}^{N} \delta(\mathbf{r}_i-\mathbf{r_j})\right\rangle_i$, where ${\langle n\rangle}$ is the average number density. 
                %In practice we compute it as a 2-dimensional histogram from a list of the particle positions.
                \item \textbf{Orientational order.} The hexatic order parameter $\psi_6$ quantifies  how close to a hexagonal arrangement the nearest neighbour of a particle are. 
                We compute it as $\psi_6(\mathbf{r_i}) = \frac{1}{N_i} \sum_{j=0}^{N_i} e^{6i\theta_{ij}} $ with $\theta_{ij}$ the angle between the $N_i$ nearest neighbors. $\psi_6(\mathbf{r_i})$ is a complex number, we  define its modulus $\vert\psi_6\vert$ and argument $\theta_6 = \frac{1}{6}\rm{arg}[\psi(\mathbf{r_i})] $. 
            \end{itemize}

        \subsubsection{Power spectra \& Dispersion relations}
\begin{figure}[t]
    \includegraphics[width=\textwidth]{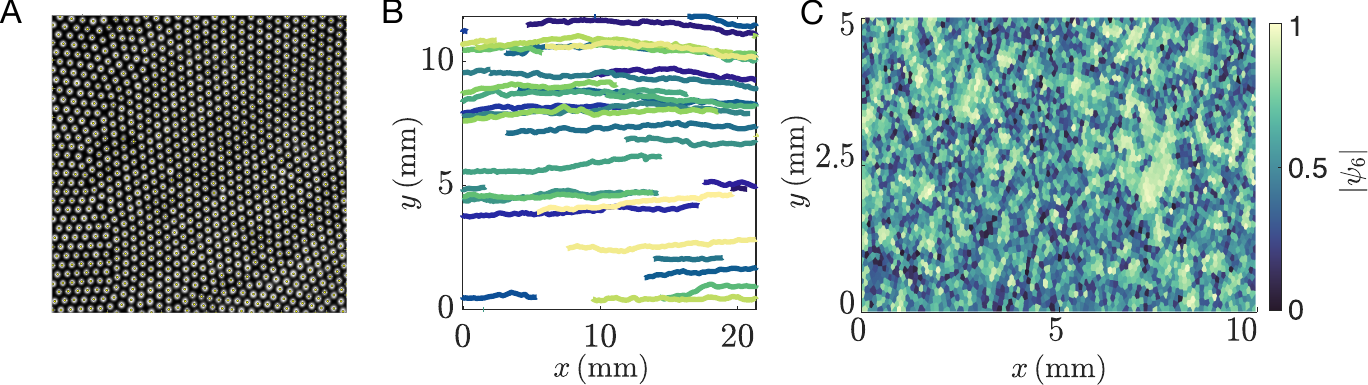}
    \caption{\textbf{Particles tracking and field reconstruction}
    {\bf A.} Particles detection :  we detect the center of mass of all droplets in the field of view using the ImageJ minima intensity detection function on contrast-enhanced and Gaussian blurred frames (the SD of the Gaussian kernel is taken roughly equal to $a/2$)
    {\bf B.} We reconstruct the drop trajectories and measure their instantaneous velocity using the MATLAB function based on the Crocker and Grier tracking algorithm.
    {\bf C.} Eulerian $\psi_6$ field in the observation region ($B=5\,\rm{(mT)}$) computed from the droplet instantaneous positions. We ensure that no large density fluctuation or packing fraction occurs in the observed region. 
    }
    \label{fig:figS2}
\end{figure}

        \paragraph{Hexatic order parameter waves}
           To define a hexatic order parameter field $\psi_6(\mathbf r,t)$ (and $\theta_6(\mathbf r,t)$),
           we interpolate the value of $\psi_6$ evaluated in each Voronoï cell on a square lattice, see Figure~\ref{fig:figS2}C. 
            The spectral power density of the Eulerian fields $f(\mathbf{r},t)=\psi_6(\mathbf{r}, t) - \overline{\psi_6}$ is given by
            \begin{equation}
            \mathcal{P}_6(\mathbf{q},\omega) = \vert\Tilde{f}(\mathbf{q},\omega)\vert^2
            \label{Pxi6}
            \end{equation}
            where, $\overline{\psi_6}$ is the space and time average of the hexatic order parameter, and $\Tilde{f}(\mathbf{q},\omega)$ is the spatio-temporal Fourier transform of  $f(\mathbf{r},t)$ in the laboratory  frame.
            In practice, we compute this spectrum using the Matlab algorithm for three-dimensional fast Fourier transform.
             We note that we keep only the values above the crystalline threshold (namely $\psi_6>0.65$) to compute all spectra, values below this threshold are set to $0$. 
            In all the experiments, the typical resolutions are $\delta q = 6.10^{-4}\,\rm{\mu m^{-1}}$ in the spacial domain, and  $\delta \omega = 0.05\,\rm{s^{-1}}$ in the temporal domain. 

            We outline the procedure for extracting the dispersion relation from the raw power spectra of $\psi_6$. First, we reduce the Fourier space to two dimensions by considering only wave-vectors in the direction of the mean flow ($q_y=0$). Next, we transition to the comoving  frame by applying a linear transformation $M = \left( \begin{smallmatrix} 1 & 0 \\ -U & 1 \end{smallmatrix} \right)$ to the spectrum such that $\omega'(q_x)=\omega(q_x)-Uq_x$. The transformed image is then linearly interpolated onto a rectangular grid and smoothed for each frequency using a moving average. Finally, the dispersion relation is extracted by identifying the maximum value of the spectrum at each frequency for every wave-vector.
                  
    \section{Propagation of density waves and dipolar hydrodynamic interactions}
\begin{figure}[t]
     \includegraphics[width=\textwidth]{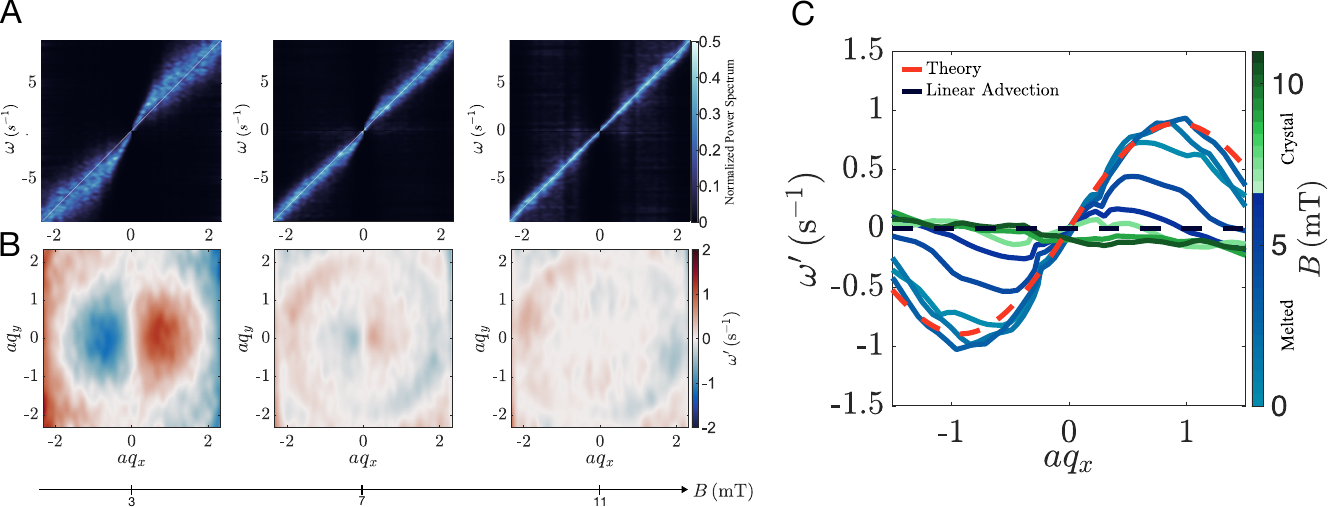}
        \caption{
        \textbf{Sound waves and dispersion relations}
        {\bf A.} Power spectra $\mathcal P_{\rho}(q_x,q_y=0,\omega)$ plotted for different values of $B$.
        In the liquid phase dispersive density waves propagate through the emulsion. Deep in the crystal phase, the density fluctuations are merely advected with the crystal.
        {\bf B.} Dispersion relations on the density waves in the frame moving with the center of mass of the emulsion ($\omega'=\omega-Uq_x$). Same values of $B$ as in {\bf A}.
        {\bf C.} Cut of the dispersion relation in the $q_y=0$ plane and comparison to theory~\cite{Desreumaux_2013}.
        }
        \label{fig:density_wave}
    \end{figure}

    In this section we provide additional evidence  that the hydrodynamic couplings between the driven droplets are well captured by the standard dipolar interactions extensively studied in e.g.~\cite{Beatus_2006,Desreumaux_2013,Beatus2017,Saeed2023}.
    Driven isotropic emulsions have been consistently shown to support sound waves despite the overdamped dynamics of all the individual particles~\cite{Beatus_2006,Desreumaux_2013}.
    The dispersion relation of the sound waves provides a direct insight into the range and symmetry of the underlying hydrodynamic interactions. 
    Following the same method as for the hexatic order parameter (see \eqref{Pxi6} and Figure 3A in the main text), we compute the fluctuation spectrum of the density fluctuations 
    \begin{equation}
            \mathcal{P}_\rho(\mathbf{q},\omega) = \vert\Tilde{\rho}(\mathbf{q},\omega)\vert^2,
            \label{Prho}
            \end{equation}
    where $\Tilde{\rho}(\mathbf{q},\omega)$ is the Fourier spectrum of the number density field $\rho(\mathbf r,t)$.
    In Figure~\ref{fig:density_wave}A, we plot the power spectra evaluated at $q_y=0$ for different values of the $B$ field. 
    In agreement with earlier results, we find that the spectra are peaked on a nontrival curve in  the $(q_x,\omega)$ plane, which reveals the propagation of sound modes.
    Increasing the value of  $B$, not surprisingly, we find that the density waves are progressively suppressed. Ultimately, we observe a simple dynamics where the  density field hardly fluctuates, and is merely advected at the mean crystal speed.  The lively dynamics of the domain walls discussed in the main text does not translate in strong density fluctuations.
    
    Locating the maxima of $\mathcal P_\rho$, and repeating the same measurement for all values of $q_y$, we can construct the dispersion relation $\omega=f(q_x,q_y)$ showed in Figure~\ref{fig:density_wave}B. 
    In the liquid phase where no crystalline order prevails, the dispersion relations are identical to that reported in Eq. 6 in~\cite{Desreumaux_2013}.
   In Figure~\ref{fig:density_wave}C, we  compare  our measurements to the theory of~\cite{Desreumaux_2013}.
   We find that over a range of $B$ fields, in the melted regime, the dispersion relation of the density waves are quantitatively captured by our model where the only interactions between the droplets are hydrodynamic dipoles and steric repulsion.
   This agreement confirms the relevance of the dipolar interactions to account for the dynamics observed in our experiments.

    \section{Robustness of our experimental results to changes in the area fraction}
\begin{figure}[h!]
    \includegraphics[width=\textwidth]{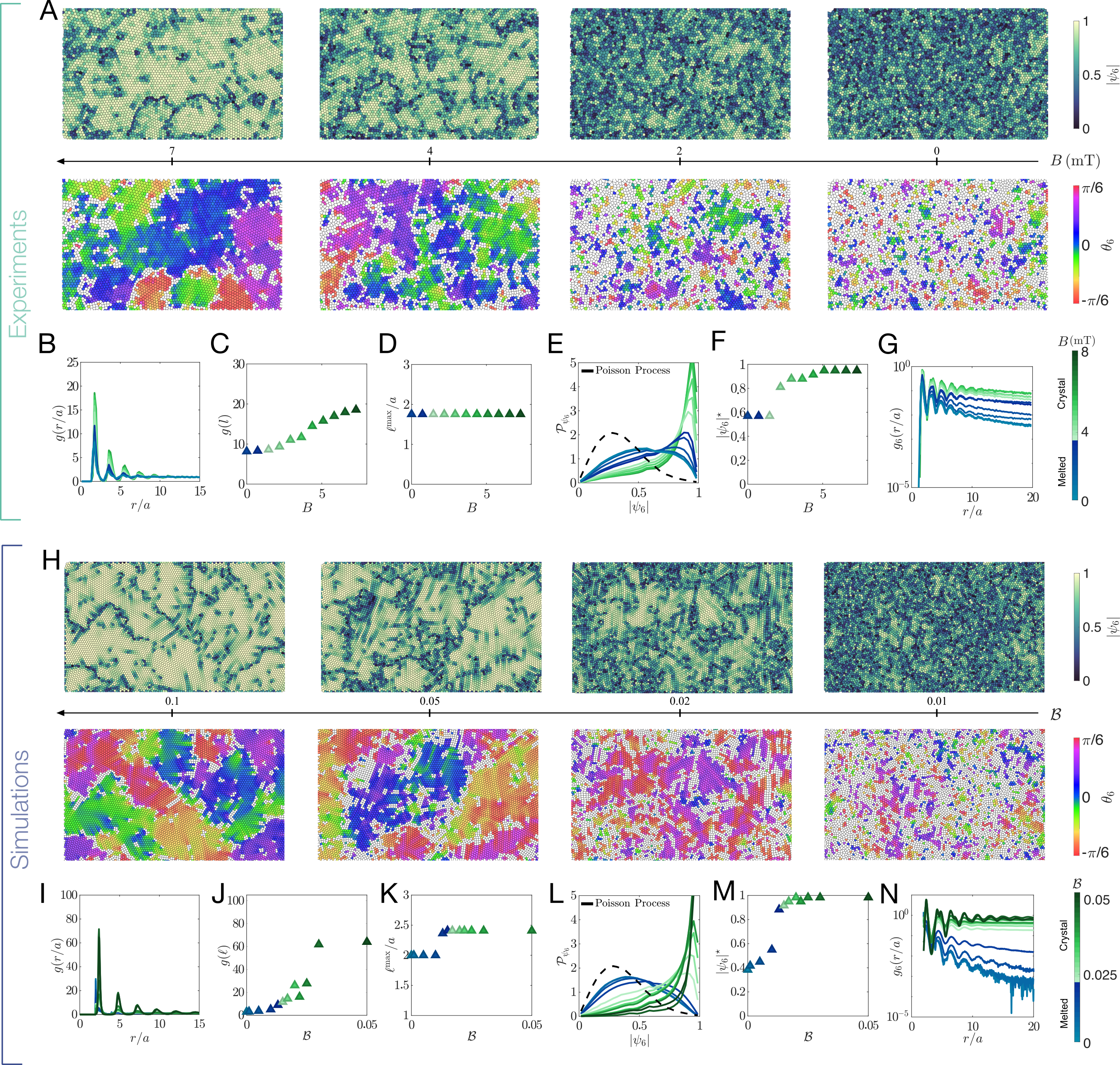}
    \caption{\textbf{Hydrodynamic melting ($\phi=0.61$)}
    {\bf A.} Maps of $\vert\psi_6\vert$ and $\theta_6$ for four different values of $B$. Packing fraction  $\phi_1=0.61$. We observe a  melting phenomenology similar to that  reported for $\phi_2=0.33$ in the main text.
    {\bf B.} Pair correlation functions $g(r/a)$ measured along the direction defined by the local value of $\theta_6$. $g(r/a)$ is plotted for different values of the applied magnetic field. 
    The inner structure of the crystal clusters hardly evolves as $B$ decreases. 
   {\bf C.} Variations of the amplitude of the first crystal peak $g(\ell)$ with $B$. 
   {\bf D.} The position of the first peak of $g(r/a)$ ($\ell$) does not vary with $B$. The structure of the Wigner crystals does not change in the bulk as $B$ decreases.
   {\bf E.} Distribution of the magnitude of the local and instantaneous hexatic field $\psi_{6}(\mathbf r,t)$. Dashed line same distribution computed for a Random set of points (2D Poissonian process).
   {\bf F.} Evolution of the position of the maximum of the $\psi_6$ distribution with the magnitude of the repulsive interactions. The position of the $\psi_6$ maximum clearly signals the melting point of the polycrystal structure.
   {\bf G.} Spatial correlations of $\psi_6$ 
   {\bf (H-N).} Comparison to numerical simulations
    }
    \label{fig:figS3}
\end{figure}

    As alluded to in the main text we have conducted a second series of experiments where the packing fraction is $\phi=0.61$, for droplets having a slightly different radius $a=47\,\rm{\mu m}$. 
    None of our main findings are qualitatively different from that reported in the main text.
    Figure~\ref{fig:figS3} reports the evolution of the translational and orientational order parameters as $B$ is varied. All our measurements confirm the robustness of the melting scenario. 
    Under the action of non-reciprocal forces, crystals melts from their grain boundaries whose the size continuously shrink as $B$ decreases.
    Of course as the packing fraction increases, the contact interactions play a more significant role and stabilizes the crystal phase. 
    Even when the $B$ field is vanishingly small, the contact interactions stabilizes the ordering of minute crystallites.

%%%%%%%%%%%%%%%%%%%%%%%%%%%%%%%%%%%%%%%%%%%%%%%%%%%%%%%%%
\section{Driven emulsion crystal: Model and numerical simulations}
\subsection{Model}
%Overdamped dynamics. No need for advection.
We detail the minimal model introduced in the main text. We consider a collection of $N$ particles that evolves according to an overdamped dynamics under the action of pairwise-additive interactions that depend only on the (vector) distance between droplets. The position $\RR_i$ of a the $i^{\rm th}$ droplet evolves in time as
\begin{align}\label{smeq:overdamped}
\zeta \dot \RR_i &= \sum_{j\neq i} \FF(\RR_i - \RR_j),
\end{align}
where $\zeta$ is a friction coefficient. The forces decompose as
\begin{align}
\FF(\rr) &= \FF_\mathrm{core}(\rr) + \FF_\mathrm{mag}(\rr) + \FF_\mathrm{hydro}(\rr),
\end{align}
with $\FF_\mathrm{core}$, $\FF_\mathrm{mag}$ and $\FF_\mathrm{hydro}$ being respectively the hard-core, magnetic and hydrodynamic forces. 
We detail them below.
Note that any advection term at velocity $\vec{V}_0$ in \eqref{smeq:overdamped} could be eliminated by a change of coordinates $\rr\mapsto\rr-\vec{V}_0t$. The mean advection plays no role in our modeling and simulations we ignore it in all that follows.

\paragraph{Hard-core repulsion.} The droplets cannot interpenetrate one another. We implement this constraint with a strongly repulsive potential Weeks-Chandler-Andersen (WCA) potential 
\begin{equation}
    \phi_\mathrm{WCA}(r) =
    \begin{cases}
    4\tilde \epsilon \left[\left(\frac{r_0}{r}\right)^{12} - \left(\frac{r_0}{r}\right)^{6}\right] + \tilde \epsilon & \mbox{if $r \leq 2^{1/6} r_0$} \\
    0 & \mbox{otherwise}
    \end{cases},
\end{equation}
with $r_0 = 2\tilde a/2^{1/6}$ with $\tilde a$ the radius of the droplets. The resulting repulsion force is simply given by $\FF_\mathrm{core}(\rr) = -\nabla \phi_\mathrm{WCA}(\rr)$.\\
%Not realistic but modeling.
%with $2a = 2^{1/6} r_0$.

\paragraph{Magnetic interactions.}
In the presence of an external magnetic field, the droplets behave as  magnetic dipoles pointing in the $z$ direction and repel one another with a repulsive dipole-dipole potential
%\FF_\mathrm{mag}(\rr) = -\nabla \phi_\mathrm{mag}(\rr)$
\begin{equation}
    \phi_\mathrm{mag}(r) = \frac{\tilde {\mathcal B}}{r^3}.
\end{equation}
The magnetic force is $\FF_\mathrm{mag}(\rr) = -\nabla \phi_\mathrm{mag}(\rr)$.
In the experiments $\tilde {\mathcal B} = \frac{\mu_0(1+\chi) m^2}{4\pi}$ with $m = \frac{\chi v B}{\mu_0}$ the magnetic moment of a droplet of volume $v$, $\chi$ the magnetic susceptibility and $m$ the magnetic dipole strength. Ce are not interested in the long-range effects of the magnetic interactions:
in the numerical simulations, a cut-off is introduced at a distance corresponding to half the box height.\\

\paragraph{Hydrodynamic interactions.} 
In a Hele-Shaw cell, the 2D velocity field $\uu(\rr)$ of a viscous fluid obeys  Darcy's law $\uu(\rr) = -\kappa \nabla P(\rr)$ with $P(\rr)$ the pressure field and $\kappa$ the permeability of the channel. The fluid is incompressible, the conituinty equations then reduces to $\nabla\cdot\uu = 0$. 
When an intruder, e.g. a squeezed droplet, moves at a velocity different from the fluid's, it induces long range perturbations on the flow. 
In the far field limit the flow perturbation induced by an finite-size introder at the orgin is well approximated as a source dipole singularity:  $\nabla\cdot\uu = -\tilde\ssigma\cdot\nabla\delta(\rr)$ with $\delta(\rr)$ the Dirac function.
The dipole strength $\tilde\ssigma$  can be expressed in terms of the area of the intruder and the velocity difference. 
This modeling can be understood as follows. To allow the intruder to move some fluid matter must be removed downstream and injected upstream hence forming a source dipole.
If another particle is placed at position $\RR$ inside this flow perturbation, it is  advected at a velocity $\zeta^{-1}\FF_\mathrm{hydro}(\RR) = \mu \uu(\RR)$ where $\mu$ is the dimensionless mobility of the particle, see e.g. Ref.~\cite{Desreumaux_2013} for more details.

Lastly, the hydrodynamic force between two droplets satisfies the equations
\begin{align}
    \FF_\mathrm{hydro}(\rr) &= - \nabla \phi_\mathrm{hydro}(\rr) \\
    \nabla\cdot \FF_\mathrm{hydro}(\rr) &= -\ssigma\cdot\nabla\delta(\rr),
\end{align}
where $\phi_\mathrm{hydro}$ is the hydrodynamic potential proportional to $P$, and $\ssigma$ is a dipole strength proportional to $\tilde\ssigma$. 
Together, these two equations imply that $\phi_{\rm hydro}$ obeys a Poisson equation,
\begin{align} \label{smeq:poisson}
    \nabla^2 \phi_\mathrm{hydro}(\rr) = q(\rr)
\end{align}
with $q(\rr) = \ssigma\cdot\nabla\delta(\rr)$. The Poisson equation is easily solved from the Green function of the 2D Laplacian:
\begin{equation}
    \phi_\mathrm{hydro}(\rr) = \ssigma\cdot\nabla\left(\frac{1}{2\pi} \ln  r\right) = \frac{1}{2\pi} \frac{\ssigma\cdot\rr}{r^2}.
\end{equation}
We stress that the hydrodynamic potential has the unusual symmetry $\nabla\phi_\mathrm{hydro}(-\rr)=-\nabla \phi_\mathrm{hydro}(\rr)$ which induces forces that do not obeys the action-reaction principle as $\FF_\mathrm{hydro}(-\rr) = \FF_\mathrm{hydro}(\rr)$.
The hydrodynamic interactions are non-reciprocal.

\subsection{Dimensionless parameters}
Considering \eqref{smeq:overdamped}, we have six parameters and three units: energy $[E]$, length $[L]$ and time $[T]$. This enables us to set three parameters to unity in the numerical simulations, or alternatively to construct three dimensionless numbers.
%: the packing fraction $\phi$, the magnetic to hydrodynamic ratio $A$ and the scaled WCA strength $\epsilon$.
%
The six dimensional parameters are: the number density $\rho$ (number of particles per surface area) with unit $[L^{-2}]$, the particle diameter $\tilde a$ with unit $[L]$, the strength of the WCA potential $\tilde \epsilon$ with unit $[E]$, the magnetic strength $\tilde A$ with unit $[EL^3]$, the hydrodynamic dipole $\sigma$ with unit $[EL]$ and the friction coefficient $\zeta$ with unit $[EL^{-2}T]$. Let us deal successively with length, energy and time scales.

\begin{itemize}
    \item We set $\rho = 1$: the typical length scale is set to $\ell^\ast = \rho^{-1/2}$. This gives us a first dimensionless number as the scaled particle radius $a = \rho^{1/2} \tilde a$, or alternatively the packing fraction $\phi = \pi \rho \tilde a^2 = \pi a^2$.
    \item We set $\sigma = 1$: the typical energy scale is set to $E^\ast = \rho^{1/2} \sigma$. This gives us two dimensionless numbers: the crucial ratio between magnetic and hydrodynamic forces ${\mathcal B} = \rho {\mathcal B}/\sigma$, and the  scaled WCA strength $\epsilon = \rho^{1/2}\tilde \epsilon /\sigma$.
    \item We set $\zeta = 1$: the timescale of the simulations is $t^\ast = \zeta / [\rho^{3/2} \sigma]$.
\end{itemize}

At the end of day, the three dimensionless parameters are: the packing fraction $\phi$, the magnetic to hydrodynamic ratio ${\mathcal B}$ and the scaled WCA strength $\epsilon$. The most important one is $A$ and we study systematically its influence on the system. As for the packing fraction, we provide data for two values: dilute system $\phi = 0.36$ ($a=0.34$) and dense system $\phi = 0.61$ ($a=0.44$). The strength of the WCA interactions is not expected to play any role in the physics of the system and we set $\epsilon = 1$ in all the simulations.

\subsection{Details of the numerical simulations}
The simulations are performed in a periodic box of size $L_x\times L_y$. $N$ particles are placed in this box, so that the density $\rho = N / (L_x L_y) = 1$. Simulations snapshots correspond to $L_x = 128$, $L_y=64$ and $N=8192$; and averaged observables to $L_x = L_y = 64$ and $N=4096$. The particles are initially placed at random and harmonic repulsive interactions are implemented to relax the system to a state where particles do not overlap (that is to say, they are at a distance larger than $2a$).

Starting from this initial condition, the forces $\FF_i(t)$ at a given time $t$ are computed as detailed below, and the positions $\RR_i(t)$ are updated according to an explicit Euler scheme: $\RR_i(t+\Delta t) = \RR_i(t) + \Delta t \FF_i(t)$.
We used a timestep $\Delta t = 0.002$ and perform up to $6\cdot 10^6$ iterations (final time $t_\mathrm{max} = 1200$). The first $2\cdot 10^6$ iterations ($t=400$) lead the system to its steady state, and the observables related to this steady state are computed by averaging over the $4\cdot 10^6$ remaining iterations.

The forces are computed as follow. The WCA interactions involve only particles touching one another. The magnetic interactions (that decay as $1/r^3$) are cut-off at a distance $L_y/2$. And the long-range hydrodynamic interactions are carefully implemented using a modified Ewald algorithm~\cite{FrenkelSmit} which we introduce below.

\subsection{Ewald algorithm for the hydrodynamic forces}
We consider $N$ dipoles $\ssigma$ at positions $\RR_i$ in a periodic box of size $(L_x, Ly)$. We call $\SSS$ the lattice vectors corresponding to this periodic box ($S_x = \frac{2\pi}{L_x} n_x$, $S_y = \frac{2\pi}{L_y} n_y$ with integers $n_x$ and $n_y$). The charge density that satisfies the Poission equation \eqref{smeq:poisson} is
\begin{equation}
q(\rr) = \sum_{i=1}^N \sum_{\SSS} \ssigma \cdot\nabla \delta\left(\rr-(\RR_i + \SSS)\right).
\end{equation}
The solution is
\begin{equation}
    \phi(\rr)  = \frac{1}{2\pi} \sum_{i=1}^N \sum_{\SSS}
    \frac{\ssigma\cdot\left(\rr-(\RR_i + \SSS)\right)}{\left(\rr-(\RR_i + \SSS)\right)^2}.
\end{equation}
In this section, we remove the subscript `$\mathrm{hydro}$' for the sake of clarity.
The hydrodynamic force acting on particle $i$ is $\FF_i=-\nabla\phi(\RR_i)$:
 \begin{equation} \label{smeq:ew_force0}
  \FF_i = \frac{1}{2\pi} \sum_{(j, \SSS) \neq (i, 0)} 
\left[\frac{2 \hat\RR_{i,j,\SSS}\hat\RR_{i,j,\SSS}-1}{\RR_{i,j,\SSS}^2} 
\right]\cdot \ssigma,
 \end{equation}
 with $\RR_{i,j,\SSS} = \RR_i - \RR_j - \SSS$, $\hat \RR_{i,j,\SSS} = \RR_{i,j,\SSS} / |\RR_{i,j,\SSS}|$ and the term with both $j=i$ and $\SSS=0$ is excluded from the sum.

The direct summation converges very slowly and is unfit for numerical simulations. We therefore adapt the Ewald summation~\cite{FrenkelSmit}, initially developed for electric charges and reciprocal electric interactions.\\

\paragraph{Decomposition.} We decompose the density into a screened part, that we compute in real space, and a Gaussian part, that we compute in Fourier space:
\begin{equation}
q(\rr) = q_R(\rr) + q_F(\rr)
\end{equation}
with
\begin{align}
q_R(\rr) &= \sum_{i=1}^N \sum_\SSS \ssigma\cdot\nabla\left[
\delta\left(\rr-(\RR_i + \SSS)\right) - \frac{\alpha^2}{\pi}
e^{-\alpha^2\left(\rr - (\RR_i+\SSS)\right)^2} \right], \\
q_F(\rr) &= \sum_{i=1}^N \sum_\SSS \ssigma\cdot\nabla\left[ \frac{\alpha^2}{\pi}
e^{-\alpha^2\left(\rr - (\RR_i+\SSS)\right)^2} \right].
\end{align}
The parameter $\alpha$ is free and is adjusted to obtain the fastest computation. We now detail how we compute the Fourier-space, and the real-space components of the hydrodynamic forces.\\

\paragraph{Fourier space component.}
We move to Fourier space ($\GG$: vector in reciprocal space, $|\mcV|$: volume of the unit cell) and decompose $q_F$ into a Fourier series
\begin{equation}
q_F(\rr) = \frac{1}{|\mcV|} \sum_\GG e^{i\GG\cdot\RR} \tilde q_F(\GG).
\end{equation}
Using $\nabla\phi_F(\rr) = q_F(\rr)$, the computation leads to
\begin{align}
\tilde q_F(\GG) &= \int_\mcV d\rr\, e^{-i\kk\cdot\GG} q_F(\rr)
= \int_{\mathbb{R}^2} d\rr\, e^{-i\kk\cdot\GG} \sum_{i=1}^N \ssigma\cdot\nabla\left[ \frac{\alpha^2}{\pi}
e^{-\alpha^2\left(\rr - \RR_i\right)^2} \right]
= (i\GG\cdot\ssigma) e^{-\frac{\GG^2}{4\alpha^2}} \sum_{j=1}^N  e^{-\GG\cdot\RR_j} \\
\tilde \phi_F(\GG) &= -\frac{1}{\GG^2} (i\GG\cdot\ssigma) e^{-\frac{\GG^2}{4\alpha^2}} \sum_{j=1}^N  e^{-\GG\cdot\RR_j}.
\end{align}
%We call $|\mcV|$ the volume of the unit cell and $\GG$ the reciprocal lattice vectors.
The real-space expression of the potential, assuming its average vanishes, is
\begin{align}
\phi_F(\rr) &= \frac{1}{|\mcV|} \sum_\GG e^{i\GG\cdot\RR} \tilde \phi_F(\GG) 
=  \frac{1}{|\mcV|} \sum_{\GG\neq\vec 0} \frac{-i\GG\cdot\ssigma}{\GG^2} e^{-\frac{\GG^2}{4\alpha^2}} \sum_{j=1}^N  e^{i\GG\cdot(\rr - \RR_j)} \\
&= \frac{1}{|\mcV|} \sum_{\GG\neq\vec 0} \frac{\GG\cdot\ssigma}{\GG^2} e^{-\frac{\GG^2}{4\alpha^2}}\sum_{j=1}^N  \sin[\GG\cdot(\rr - \RR_j)]
\end{align}
And the force on particle $i$, $\FF_{F,i} = -\nabla\Phi_F(\RR_i)$ excluding the term $j=i$, is
%\begin{equation}
%\ff_{F, i} =  \frac{-1}{|\mcV|} \sum_{\GG\neq\vec 0} \left\{ \frac{\GG\cdot\ssigma}{\GG^2} e^{-\frac{\GG^2}{4\alpha^2}} \sum_{j\neq i}  \cos[\GG\cdot(\RR_i - \RR_j)]\right\} \GG
%\end{equation}
\begin{equation} \label{smeq:ew_fourier}
\FF_{F, i} =  \frac{-1}{|\mcV|} \left\{
(N-1)\frac{\ssigma}{2} + \sum_{\GG\neq\vec 0}  \left(\sum_{j\neq i}  \cos[\GG\cdot(\RR_i - \RR_j)]\right)  e^{-\frac{\GG^2}{4\alpha^2}} \frac{(\GG\cdot\ssigma) \GG}{\GG^2}\right\}.
\end{equation}
This sum converges exponentially fast. 
One subtlety is that there is a component for $0$ wave-number. It comes from the relation
$\frac{(\kk\cdot\ssigma)\kk}{\kk^2} \sim \frac{\ssigma}{2}$ as $\kk\to\vec 0$ ``in a isotropic way'' that can be obtained by inverting the Fourier transform and computing the integral over space with integration first perform over the angular coordinate.

Another important remark is that the sum over $j$ can computed from the structure factor $S(\GG)$,
\begin{gather}
    \sum_{j\neq i}  \cos[\GG\cdot(\RR_i - \RR_j)]
    = \mathrm{Re}\left[S(\GG)\right] \cos(\GG\cdot\RR_i) + \mathrm{Im}\left[S(\GG)\right] \sin(\GG\cdot\RR_i) - 1, \\
    S(\GG) = \sum_j e^{\GG\cdot\RR_j}.
\end{gather}
The computation of the $N$ forces is therefore done in linear time and not in quadratic time, and the constant factors can be precomputed.\\

\paragraph{Real space component.}
To compute the real-space contribution, we first investigate the case of a single particle in infinite space,
\begin{align}
q_R^{(1)}(\rr) &= \ssigma\cdot\nabla\left[
\delta\left(\rr\right) - \frac{\alpha^2}{\pi}
e^{-\alpha^2\rr^2} \right], \\
\tilde q_R^{(1)}(\kk) &= i\kk\cdot\ssigma \left[
1 - 
e^{-\frac{\kk^2}{4\alpha^2}} \right], \\
\tilde \phi_R^{(1)}(\kk) &= \frac{-i\kk\cdot\ssigma}{\kk^2} \left[
1 - 
e^{-\frac{\kk^2}{4\alpha^2}} \right].
\end{align}
Inverting the Fourier transform leads to
\begin{align}
\phi_R^{(1)}(\rr) &= \frac{-i\sigma}{(2\pi)^2} \int_0^\infty dk \left(1-e^{-\frac{k^2}{4\alpha^2}}\right) \int_0^{2\pi} d\varphi e^{ikr\cos(\varphi-\theta)} \cos(\varphi-\alpha) %\\
%&= \frac{\ssigma\cdot\hat \rr}{2\pi|\rr|} \int_0^\infty dq \left(1-e^{-\frac{q^2}{4\alpha^2 r^2}}\right) J_1(q) \\
%&
= \frac{\ssigma\cdot\hat \rr}{2\pi r} e^{-\alpha^2 \rr^2}, \\
-\nabla \phi_R^{(1)}(\rr) &=
%\frac{e^{-\alpha^2 \rr^2}}{2\pi|\rr|^2} \left[2(1+\alpha^2)(\hat\rr\hat\rr) - \mbI\right]\cdot \ssigma
\frac{e^{-\alpha^2 \rr^2}}{2\pi} \left[\frac{2(\ssigma\cdot\hat\rr)\hat\rr - \ssigma}{\rr^2} + 2\alpha^2(\ssigma\cdot\hat\rr)\hat\rr\right],
\end{align}
with $\hat\rr = \rr/r$.
We can now sum all the contributions and obtain
\begin{align}
\phi_R(\rr) &= \frac{1}{2\pi} \sum_{i=1}^N \sum_\SSS \frac{\ssigma\cdot(\rr-\RR_i-\SSS)}{(\rr-\RR_i-\SSS)^2} e^{-\alpha (\rr-\RR_i-\SSS)^2}, \\
\FF_{R, i} &= \sum_{j\neq i} \sum_\SSS
\frac{e^{-\alpha^2 \RR_{i,j,\SSS}^2}}{2\pi}
\left[\frac{2 \hat\RR_{i,j,\SSS}\hat\RR_{i,j,\SSS}-1}{\RR_{i,j,\SSS}^2} 
+ 2\alpha^2 \hat\RR_{i,j,\SSS}\hat\RR_{i,j,\SSS} \right]\cdot \ssigma. \label{smeq:ew_real}
\end{align}
The sum converges exponentially fast.

\paragraph{Self-interaction.} In \eqref{smeq:ew_fourier} and \eqref{smeq:ew_real}, we excluded the term $j=i$ from the summation as a droplet is not advected by its own flow.  However, in periodic space a droplet does interact with its images. This is irrelevant for reciprocal forces ($\FF(-\rr) = -\FF(\rr)$) since the contributions from the images compensate. But in our non-reciprocal case ($\FF(-\rr) = \FF(\rr)$), the interaction of a droplet with its images, that we call ``self-interaction'', can be non-zero. The self-interaction force $\FF_\mathrm{self}$ is the same for all particles and reads
\begin{equation} \label{smeq:ew_self}
    \FF_\mathrm{self} = \frac{1}{2\pi}\sum_{\SSS\neq 0}
\left[\frac{2 \hat\SSS\hat\SSS-1}{\SSS^2} \right]\cdot \ssigma,
\end{equation}
where the sum is on non-zero reciprocal space vector and $\hat \SSS = \SSS / |\SSS|$. 
This sum is not absolutely convergent: we consider the limit corresponding to a summation over the same number of boxes in the $x$ and $y$ directions.

\paragraph{Summary.} At the end of the day, the slowly converging sum from \eqref{smeq:ew_force0}, is replaced by
\begin{equation}
    \FF_i = \FF_{f,i} + \FF_{r, i} + \FF_\mathrm{self},
\end{equation}
with $\FF_{r,f}$ given by \eqref{smeq:ew_fourier} which converges exponentially fast in Fourier space, $\FF_{r,i}$ given by \eqref{smeq:ew_real} which converges exponentially fast in real space, and $\FF_\mathrm{self}$ given by \eqref{smeq:ew_self} that is computed only once.
The parameter $\alpha$ is free and is optimized such that the trade-off between computations in Fourier and real spaces is as fast as possible.

\section{Non-reciprocal interactions propel dislocations}
In this section we generalize the results presented in~\cite{Poncet2022}. Both in our experiments and numerical simulations, we observe that the crystal defects  propel in the upstream direction, see Fig.~3.
We explain our observations based on kinematics and symmetry arguments in the discussion section of the main text.
Here, we  provide a more technical and accurate explanation for the propulsion of  dislocations powered by non-reciprocal interactions. 
This section is organized as follows: (i) we discuss the stability of crystals challenged by non-reciprocal forces.  In particular, we show that  hydrodynamic interaction arising from Darcy's flows cannot lead to a linear instability of driven crystals.
(ii) We provide a general expression for $\bf F^{\rm NR}$ in the case of short-ranged non-reciprocal interactions, and discuss its relation to the early model introduced in~\cite{Lahiri_1997}. 
 (iii) We focus on the long-range  hydrodynamic interactions relevant to our experiments, we compute $\mathbf F^{\rm NR}$ and $\mathbf u^{\rm NR}$. 
 (iv) We put our results in the context of the Peach-Koehler force formalism~\cite{Oswald}.
 (v) We explain how dislocations glide and split in response to $\mathbf F^{\rm NR}$.
 (vi) Lastly, we discuss the impact of our results on the dynamics of grain-boundaries.

\subsection{Can non-reciprocal forces destabilize driven lattices?}
\subsubsection{General case and potential flows}
For the sake of clarity, unlike in Section 4 and  Eq. 1 in the main text,  we consider here the overdamped dynamics of a hexagonal lattice of particles interacting only via non-reciprocal forces:
\begin{equation}
\zeta\partial_t \mathbf{R}_i = \sum_{j \neq i} \mathbf{F}(\mathbf{R}_i - \mathbf{R}_j), 
\end{equation}
and by definition  $\mathbf{F}(-\mathbf{r}) = \mathbf{F}(\mathbf{r})$. 
For instance $\mathbf F$ can represent $\mathbf F_{\rm hydro}$ introduced in Eq. 1 in the main text. 
However, we henceforth keep the discussion more general and specify the form and origin of the non-reciprocal force only when needed.

To study the stability of the arrangement of particles, we define the displacement field $\uu(\rr)$ and its Fourier transform $\tilde \uu(\qq)$. The conventions are
	\begin{align}
		\tilde \uu(\qq) &= \int d\rr\, e^{-i\qq\cdot\rr} \uu(\rr), &
		\uu(\rr) &= \frac{1}{(2\pi)^2} \int d\qq\, e^{+i\qq\cdot\rr} \tilde \uu(\qq).
	\end{align}
The resulting linear dynamics is then defined by the stability matrix $M(q)$ as
	\begin{align}
		\zeta\partial_t \tilde\uu(\qq) &= M(\qq)\cdot \tilde\uu(\qq)  \label{eq:uq} \\
		M(\qq) &= i\sum_{j\neq 0} \sin(\qq\cdot \rr_j) \nabla\FF(\rr_j) \label{eq:Mq}
	\end{align}
	where $\rr_j$ (norm $r_j$ and angle $\theta_j$) are the positions of the particles, see~\cite{Poncet2022}. We used the fact that $\FF(-\rr)=\FF(\rr)$. 
    We note that the inverse Fourier transform of the right-hand side of \eqref{eq:uq} is the non-reciprocal force field $\FF^{\rm NR}$.
    As discussed in~\cite{Poncet2022}, without additional constraints on $\mathbf F$, the dynamics is either marginally stable or unstable. 
    However when the non-reciprocal forces arise from 2D potential fluid flows, such as viscous Darcy's flows,  $\mathbf F$ enjoys an addition property. 
    $\bf F$ can be either written in term of a potential $\mathbf F=\bm \nabla \phi$, or a stream function $\mathbf F=\bm \nabla^\perp \psi$, so that:
    \begin{align}
		\nabla\FF &= \begin{pmatrix}
			\frac{\partial F_x}{\partial x} & \frac{\partial F_x}{\partial y} \\ \frac{\partial F_y}{\partial x} & \frac{\partial F_y}{\partial y}
		\end{pmatrix} = \begin{pmatrix}
			\frac{\partial^2 \psi}{\partial x\partial y} & \frac{\partial^2 \phi}{\partial x\partial y} \\ \frac{\partial^2 \phi}{\partial x\partial y} & -\frac{\partial^2 \psi}{\partial x\partial y}
		\end{pmatrix}
	\end{align}
    The sum in \eqref{eq:Mq} leads to a matrix of the form
	\begin{equation}
		M(\qq) = i \begin{pmatrix}
			m_1 & m_2 \\ m_2 & -m_1,
		\end{pmatrix}
	\end{equation}
	where $m_1$ and $m_2$ are real numbers. 
    The eigenvalues are $\pm i \sqrt{m_1^2+m_2^2}$ and the system is therefore marginally stable.

\subsubsection{2D lattices hydrodynamically driven through viscous fluids are marginally stable}
In the specific case of dipolar hydrodynamic interactions, $\phi(\rr)=\sigma\cos \theta/r$, and $\psi(\rr)=-\sigma\sin\theta/r$. The nonreciprocal hydrodynamic force takes the form
\begin{align}
		\FF(\rr) &=  \frac{-\sigma}{r^2}\begin{pmatrix}
			\cos(2\theta) \\ \sin(2\theta)
		\end{pmatrix}, \label{eq:fhydro}
        \end{align}
Note that, to simplify the algebra, our notation of the strength  $\sigma$ of the dipolar force differs from that used in Section 4.

        The associated Jacobian matrix reads
        \begin{align}
		\nabla\FF(\rr) &=  \frac{2\sigma}{r^3} \begin{pmatrix}
			\cos(3\theta) & \sin(3\theta) \\ \sin(3\theta) & -\cos(3\theta)
		\end{pmatrix},
	\end{align}
    and the stability matrix is given by
\begin{align}\label{eq:MqDip}
		M(\qq) &= 2i\sigma \sum_{j\neq 0} \frac{ \sin(qr\cos(\theta_j-\varphi))}{r_j^3} \begin{pmatrix}
			\cos(3\theta_j) & \sin(3\theta_j) \\ \sin(3\theta_j) & -\cos(3\theta_j).
		\end{pmatrix}
	\end{align}
We can compute the sum \eqref{eq:MqDip} numerically for an hexagonal crystal having the shape of a circle of radius $R$ and a unit lattice spacing ($\sigma=1$). 
In limit of infinitely large systems ($qR\gg 1$), we find
	\begin{align}
		m_1 &\approx -1.8 q\cos(3\varphi), & m_2 &\approx -1.8 q\sin(3\varphi),
	\end{align}
	independently of the orientation of the crystal, where $\varphi$ is the angle of the wavevector $\qq$. 
	To make sense of this scaling, we can approximate the sum \eqref{eq:Mq} by its continuum equivalent assuming a homogeneous density $\bar\rho = 2/(a^2\sqrt{3})$, where $a$ is the lattice spacing.
    This leads to 
	\begin{align}
		m_1 &= 2\sigma\bar\rho\int_0^\infty dr\, r \int_0^{2\pi} d\theta 
		\frac{\sin(qr\cos(\theta-\varphi))}{r^3}\cos(3\theta) 
        %=
		%\frac{-4\pi}{a^2\sqrt{3}} \cos(3\varphi)\int_0^\infty dr \frac{J_3(q r)}{r^2} \\ &
        = -\frac{\pi\sigma\bar\rho}{2}q \cos(3\varphi), \\
		m_2 &= 2\sigma\bar\rho\int_0^\infty dr\, r \int_0^{2\pi} d\theta 
		\frac{\sin(qr\cos(\theta-\varphi))}{r^3}\sin(3\theta)  = -\frac{\pi\sigma\bar\rho}{2} q\sin(3\varphi).
	\end{align}
	We note that $\frac{\pi\bar\rho}{2}= \frac{\pi}{\sqrt{3}} = 1.81$ in quantitative agreement with the numerical summation ($a=1$).
    This result is consistent with the earlier predictions of~\cite{Desreumaux_2013,Saeed2023}: driven hydrodynamic crystals are marginally stable. It is worth noting that the addition of elastic forces can only further stabilize the crystal dynamics, see e.g.~\cite{Tlusty_2021} for a comprehensive discussion.

\subsection{Short-range nonreciprocal interactions}\label{sec:shortrange}
\subsubsection{Computation of $F^{\rm NR}$}
When the nonreciprocal forces $\mathbf F$ are short ranged (we note $\ell$ their typical range), we can take the long wavelength limit $q\ell\ll 1$ in \eqref{eq:Mq}, which yields
	\begin{align}
		M(\qq) &\equi{q\ell\ll 1} \sum_{j\neq 0} (i\qq\cdot \rr_j) \nabla\FF(\rr_j). \label{eq:MqLow}
	\end{align}
This relation allows us to express $\mathbf F^{\rm NR}$ as a gradient expansion:
\begin{align}
F_\alpha^{\rm NR}=A_{\alpha\beta\gamma}\partial_\gamma u_\beta
\end{align}
where 
\begin{align}
A_{\alpha\beta\gamma}=\sum_j r_{j,\beta}\partial_\gamma F_\alpha(\mathbf r_j).
\end{align}
The above tensor depends on the symmetries of both the crystal structure and of the force $\mathbf F$.

To illustrate the above calculation on a specific example, we consider short range interactions having the same dipolar symmetry as the hydrodynamic dipoles:
\begin{align}
		\FF(\rr) &= f(r) \begin{pmatrix}
			\cos(2\theta) \\ \sin(2\theta)
		\end{pmatrix}, \\
		\nabla\FF(\rr) &= \begin{pmatrix}
			\cos\theta\left[\cos(2\theta) f'(r) + 4\sin^2\theta \frac{f(r)}{r}\right] &
			\sin\theta\left[\cos(2\theta) f'(r) - 4\cos^2\theta \frac{f(r)}{r}\right] \\
			\cos\theta\sin(2\theta) f'(r) - 2\cos(2\theta)\sin\theta \frac{f(r)}{r} &
			\sin(2\theta)\sin\theta f'(r) + 2\cos(2\theta)\cos\theta \frac{f(r)}{r}
		\end{pmatrix}. \label{eq:jacSR}
	\end{align}
	For instance $f(r) = -e^{-r/\ell}$. Using the six-fold symmetry of the lattice in \eqref{eq:MqLow} we find 
    \begin{align}
		M(\qq) &\equi{qb\ll 1} K\begin{pmatrix}
			iq\cos\varphi & -iq \sin\varphi \\
			iq \sin\varphi & iq\cos\phi
		\end{pmatrix} = K\begin{pmatrix}
		iq_x & -iq_y \\
		iq_y & iq_x
		\end{pmatrix}, \\
		K &= \frac{1}{4}\sum_j \left[2f(r_j) +r_jf'(r_j)\right]. \label{eq:exprK}
	\end{align}
In other words we can express $\mathbf F^{\rm NR}$ as
	\begin{equation}
		\mathbf F^{\rm NR} =
		K\begin{pmatrix} \frac{\partial u_x}{\partial x} - \frac{\partial u_y}{\partial y} \\\frac{\partial u_x}{\partial y} + \frac{\partial u_y}{\partial x} \end{pmatrix}
		\label{eq:dynamicsSR}
	\end{equation}
Two comments are in order. 
Firstly changing the range of the dipolar interactions has a strong impact on the stability of the crystal. 
The eigenvalues of $M(\mathbf q)$ are $\lambda_\pm = i Kq_x \pm Kq_y$. 
The crystal is therefore unstable as soon as $q_y\neq 0$. 
Secondly, it is worth noting that this linear dynamics was conjectured in~\cite{Lahiri_1997} to model crystals driven through a frictional medium.
The above analysis shows that this simplified model does not correctly account for the 
dynamics of 2D crystals driven through confined viscous fluids.

\subsection{Long-range dipolar interaction: $F^{\rm NR}$ and dislocation dynamics}
\subsubsection{Computation of $\mathbf F^{\rm NR}$}
 In this section we compute the non-reciprocal force field $F^{\rm NR}$ that arises from microscopic hydrodynamic interactions relevant to our experiments. 
 We recall that these microscopic forces are long ranged and have a dipolar symmetry, they take the form:
 \begin{align}
		\FF(\rr) &=  \frac{-\sigma}{r^2}\begin{pmatrix}
			\cos(2\theta) \\ \sin(2\theta)
		\end{pmatrix}. 
\end{align}
As detailed in above, the linear dynamics of the displacement field is given by
\begin{align}
		\zeta\partial_t \tilde\uu(\qq) &= M(\qq)\cdot \tilde\uu(\qq),
	\end{align}
where the stability matrix is
	\begin{equation}
		M(\qq) = -\frac{i\pi\sigma\bar\rho}{2}q \begin{pmatrix}
			\cos(3\varphi) & \sin(3\varphi) \\ \sin(3\varphi) & -\cos(3\varphi)
		\end{pmatrix}.
	\end{equation}
Equivalently, in real space the crystal dynamics then takes the form $	\partial_t \uu(\rr)=\mathbf F^{\rm NR}(\lbrace\mathbf u(\mathbf r)\rbrace)$, where $\mathbf F^{\rm NR}(\lbrace\mathbf u(\mathbf r)\rbrace)$ is a non-local function of the strain field as a result of the long-range nature of the interactions. Computing the Fourier transform of $M(\qq)$, we obtain
\begin{align} \label{eq:dyn2}
		\mathbf F^{\rm NR}(\lbrace\mathbf u(\mathbf r)\rbrace)& = \int d\rr' M(\rr-\rr')\cdot \uu(\rr'), \\
        M_{kl} &= -\bar\rho \frac{\partial F_k}{\partial r_l}, \\
        \nabla\FF &= \frac{2\sigma}{r^3} \begin{pmatrix}
		\cos(3\theta) & \sin(3\theta) \\ \sin(3\theta) & -\cos(3\theta)
		\end{pmatrix}.
	\end{align}
    At this stage it is worth noting that $\mathbf F^{\rm NR}$ is nothing else but the convolution of the microscopic dipolar force $\bf F$ with the local density fluctuation $\delta\rho\simeq -\bar{\rho}\bm \nabla\cdot \mathbf u$. This can be readily seen by writing
	\begin{align}
    \label{eq:convolution}
		F_k^{NR} &= -\bar\rho\int d\rr'\frac{\partial F_k}{\partial r_l}(\rr-\rr')u_l(\rr')
		=-\bar\rho \int d\rr' F_k(\rr')\frac{\partial u_l}{\partial r_l}(\rr-\rr'), \\
		\mathbf F^{\rm NR}(\lbrace\mathbf u(\mathbf r)\rbrace) &= -\bar\rho \int d\rr' \FF(\rr') \nabla\cdot\uu (\rr-\rr').
	\end{align}
 This result will prove very useful to address the dynamics of an isolated dislocation below.

 \subsubsection{Computation of $\dot\uu^{\rm NR}$ around an isolated dislocation}   
We now compute the velocity field  induced by the nonreciprocal forces around an isolated dislocation at the origin. It corresponds to $\dot{\mathbf u}^{\rm NR}$ defined in the main text.
We define the Burgers vector of the dislocation $\bb = b\hat\ee_\beta$ as $\oint \partial_i u_j dr_i = b_j$.
When the particle forming the lattice are elastically coupled a dislocation induces displacements that decay algebraically as
	\begin{align}
    \uu(\rr) &=\frac{b}{2\pi}\left[\theta \begin{pmatrix}\cos\beta \\ \sin\beta\end{pmatrix} 
    + \frac{1-\nu}{2} \log r \begin{pmatrix}\sin\beta \\ -\cos\beta\end{pmatrix} + \frac{1+\nu}{2}\cos(\theta-\beta)\begin{pmatrix}\sin\theta \\ -\cos\theta\end{pmatrix} \right]
	\end{align}
   where $\nu$ is the Poisson ratio. The corresponding deformations read
	\begin{align}
		\frac{\partial u_x}{\partial x} + \frac{\partial u_y}{\partial y} &=\frac{b(1-\nu)}{2\pi r}\sin(\beta-\theta), &
		\frac{\partial u_x}{\partial y} - \frac{\partial u_y}{\partial x} &= \frac{b}{\pi r}\cos(\theta-\beta),  \label{eq:disloc_dilat} \\
		\frac{\partial u_x}{\partial x} - \frac{\partial u_y}{\partial y} &= -\frac{b(1+\nu)}{2\pi r}\cos(\theta-\beta)\sin(2\theta), &
		\frac{\partial u_x}{\partial y} + \frac{\partial u_y}{\partial x} &= \frac{b(1+\nu)}{2\pi r}\cos(\theta-\beta)\cos(2\theta), \label{eq:disloc_shear}
	\end{align}
We can now use \eqref{eq:convolution} to compute $\dot\uu^{\rm NR}$. The convolution product is easily evaluated in Fourier space: $\tilde{\dot\uu}^{\rm NR}(\mathbf q)=\zeta^{-1}\tilde{\mathbf F}^{\rm NR}(\qq)=\zeta^{-1}\delta\tilde\rho(\qq) \tilde\FF(\qq)$. We find
\begin{align}
\tilde{\dot\uu}^{\rm NR}(\mathbf q)=
i\zeta^{-1}b\pi^2(1-\nu) \sigma\bar\rho \frac{\sin(\beta-\varphi)}{q}\begin{pmatrix}
			\cos(2\varphi) \\ \sin(2\varphi)
		\end{pmatrix},
\end{align}
where we have used
	\begin{align}
		\tilde\FF(\qq) &= \int_0^\infty dr\, r \int_{0}^{2\pi} d\theta e^{-iqr\cos(\theta-\varphi)} \frac{-\sigma}{r^2}\begin{pmatrix}
			\cos(2\theta) \\ \sin(2\theta)
		\end{pmatrix} = \pi \sigma \begin{pmatrix}
			\cos(2\varphi) \\ \sin(2\varphi)
		\end{pmatrix},
        \end{align}
        and
        \begin{align}
		\delta\tilde\rho(\qq) &= -\bar\rho\int_0^\infty dr\, r \int_{0}^{2\pi} d\theta e^{-iqr\cos(\theta-\varphi)} \frac{b(1-\nu)}{2\pi r}\sin(\beta-\theta)
		= ib(1-\nu)\pi\frac{\sin(\beta-\varphi)}{q}.
	\end{align}
We can then express the flow field in real space and find
	\begin{align}
		\dot\uu^{\rm NR}(\rr) &= i\zeta^{-1}b\pi^2(1-\nu) \sigma\bar\rho \frac{1}{(2\pi)^2} \int_0^\infty dq\, q \int_{0}^{2\pi} d\varphi e^{+iqr\cos(\varphi-\theta)} \frac{\sin(\beta-\varphi)}{q}\begin{pmatrix}
			\cos(2\varphi) \\ \sin(2\varphi)
		\end{pmatrix} \\
		%&= ib\pi^2(1-\nu) \sigma\bar\rho \frac{2i\pi}{(2\pi)^2} \int_0^\infty dq \left[ %J_1(qr)\sin(\beta-\theta)\begin{pmatrix}
		%	\cos(2\theta) \\ \sin(2\theta)
		%\end{pmatrix}
		%-2 \frac{J_2(qr)}{qr} \begin{pmatrix}
		%	\sin(\beta - 3\theta) \\ \cos(\beta-3\theta)
		%\end{pmatrix}
		%\right] \\
		&= \frac{\pi\zeta^{-1} b(1-\nu)\sigma \bar\rho}{2r} \cos(\beta-\theta) \begin{pmatrix}
			-\sin(2\theta) \\ \cos(2\theta)
		\end{pmatrix}.
        \label{eq:unrlr}
	\end{align}

\begin{figure}[hbtp]
    \includegraphics[width=\textwidth]{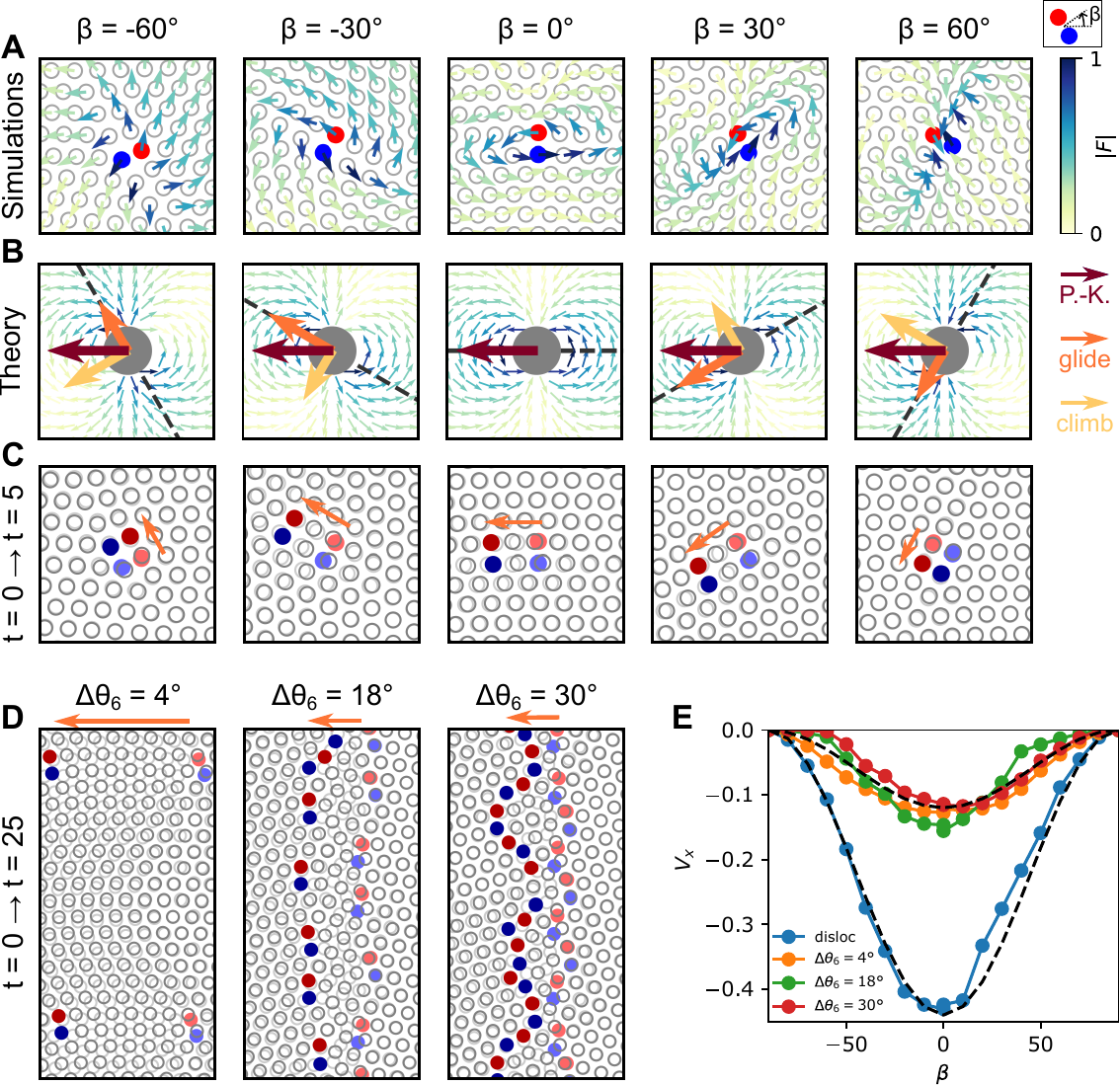}
        \caption{\textbf{Peach-Koehler force and dislocation glide.}
        {\bf A.} Dislocation with a Burgers vector at angle $\beta$ with the horizontal axis ($\beta = -60\si{\degree}, -30\si{\degree}, 0\si{\degree}, 30\si{\degree}, 60\si{\degree}$). The red (resp. blue) filled particle has 5 (resp. 7) neighbors. The arrows are the hydrodynamic forces $\FF_i$ acting on each particle (the norm is coded by the colormap).
        {\bf B.} Depiction of the theoretical force field, $F(r, \theta)$, from \eqref{eq:unrlr} with the same colormap. We see a good agreement with the microscopic forces in panel A. The dashed line is the glide line of the dislocation. The three arrows depict the effective Peach-Koehler force (\eqref{smeq:FPK3}), its glide component and its climb component (darker to lighter colors).
        {\bf C.} When the initial condition is evolved with additional strong magnetic interactions ($\mathcal{B}=0.2$), the dislocation always glide towards the left, as expected. The initial state ($t=0$) is shown with light red, blue and grey colors, while the state at $t=0$ is shown with dark colors. The arrow indicate the glide of the dislocation.
        {\bf D.} Flat grain boundaries with mismatch angles $\Delta\theta_6 = 18\si{\degree}$ and $30\si{\degree}$ shown between $t=0$ and $t=25$, with the same convention as panel C. The angle of the grain compared to the horizontal axis is $\beta = 0$.
        {\bf E.} Steady-state velocity $V_x$ along the $x$-axis for an isolated dislocation and grain boundaries with mismatches $\Delta\theta_6 = 4\si{\degree}, 18\si{\degree}$ and $30\si{\degree}$, as a function of the angle $\beta$ with the horizontal axes. The dashed lines show dependencies as $-\cos^2\beta$ predicted by our theory. 
        }
        \label{fig:disloc}
    \end{figure}

    This far-field calculation captures well the symmetries and the decay of the sum of all the microscopic forces as shown in Fig.~\ref{fig:disloc}A-B. 
    Before explaining how this force field  drives the motility and  splitting of the dislocations we can make two side comment.
    We note that point defects such as interstitials  also induce a nontrivial force field on the surrounding particles. Following the same procedure as above, we find that it has  a dipolar symmetry too.
    It is well described  by \eqref{eq:fhydro},  $\mathbf r$ being now the distance to the defect core. 
    The same results holds for vacancies for which \eqref{eq:fhydro} has to be corrected by a sign factor.

    Lastly it is worth noting that the symmetries of the flows induced by dipolar interactions around a dislocation (Eq.~\eqref{eq:unrlr}) are robust to  the interaction range.
    Considering screened  dipolar interactions (see Section \ref{sec:shortrange}), we can combine \eqref{eq:dynamicsSR} and \eqref{eq:disloc_shear} to find that they yield 
    $\dot\uu^{\rm NR}$ flows having the same symmetries as in Fig.~\ref{fig:disloc}A-B, see also~\cite{Poncet2022}.

\subsection{Effective Peach-Koehler force}

Fig.~\ref{fig:disloc}A shows that the hydrodynamic forces, if unopposed, would lead to shear and/or dilation deformations.
But as they are opposed by elastic forces, they translate into shear and dilation stresses within the crystal. 
Instead of using the kinematic argulments from the main text to account for the dislocation dynamics, we can instead  opt for the classical Peach-Koehler-force argument~\cite{Peach_1950}.  
In short, a dislocation of Burgers vector $\bb$ in an external stress field $\ssigma$ can release elastic energy by moving.
The energy gain per unit distance translates into a force commonly referred to  as the Peach-Koehler force $\FF^\mathrm{PK}$. It takes the form
\begin{equation}
    F_i^\mathrm{PK} = \epsilon_{ij} \sigma_{kj} b_k,
\end{equation}
where $\epsilon_{ij}$ is the totally antisymmetric tensor~\cite{Peach_1950,Oswald}.
This force is of purely topological origin and does not depend on the specifics of the constitutive relation that relates the stress to the strain field.

When the magnetic interactions are strong, our system is an elastic material driven by  a self-induced non-reciprocal-force field $\FF$, \eqref{eq:unrlr}.
We showed in~\cite{Poncet2022} that we can estimate the magnitude  of the effective Peach-Koehler force that arise form the  hydrodynamic forces as 
\begin{equation} \label{smeq:FPK2}
    F_i^\mathrm{PK} = \epsilon_{ij} \left.\frac{\partial F_k^{\rm NR}}{\partial r_j}\right|_\text{disloc} b_k,
\end{equation}
where $\left.\frac{\partial F^{\rm NR}_k}{\partial r_j}\right|_\text{disloc}$ defines the force difference across the dislocation, it plays the role of an effective stress acting on it.
In Ref.~\cite{Poncet2022}, we showed how to estimate this force gradient and found
\begin{equation}
    \left.\frac{\partial F_i}{\partial r_j}\right|_\text{disloc} = \frac{f_0}{b} \begin{pmatrix}
        -\sin\beta & -\cos\beta \\ \cos\beta & -\sin\beta
    \end{pmatrix}
\end{equation}
where $f_0$ is a constant having the dimension of a force, $\beta$ is the orientation of the Burgers vector.
Introducing this expression into \eqref{smeq:FPK2}, we find a  simple expression for the effective Peach-Koehler force acting on the dislocation:
\begin{equation} \label{smeq:FPK3}
    \FF^\mathrm{PK} = -f_0\hat\ee_x.
\end{equation}
It is always oriented in the direction opposite to the flow that induces the dipolar interactions, and is independent of the orientation of the Burgers vector (Fig.~\ref{fig:disloc}B).

\subsection{Gliding, climbing and splitting}
$\FF^\mathrm{PK}$ tells us that  some elastic energy is gained when the dislocation moves towards the left hand side. 
But it doesn't say whether this motion is possible and does not specifies its  kinematics either.
It is standard to distinguish the  motion of the dislocation along its Burgers vector, known as glide motion, from a motion orthogonal to the Burgers vector, know as climb motion \cite{Landau}.
While glide is easily achieved by local rearrangements around the dislocation, climb is usually forbidden microscopically. It can happen only directly with a transfer of mass, or effectively by the splitting of the dislocation in two gliding dislocations~\cite{Irvine_2013}.
We investigate the relevance of gliding and climbing scenarios in the dynamics that we observe.
To do this, we first decompose the effective Peach-Koehler force, \eqref{smeq:FPK3}, on the basis made by the normalized Burgers vector $\hat \bb=\bb/b$ and its orthogonal vector $\hat\bb^\perp$ ($\hat b^\perp_i = -\epsilon_{ij} \hat b_j$) (see also Fig.~\ref{fig:disloc}B):
\begin{align}
    \FF^\mathrm{PK} &= \FF^\mathrm{PK}_\mathrm{glide} + \FF^\mathrm{PK}_\mathrm{climb}, \\
    \FF^\mathrm{PK}_\mathrm{glide} &= (\FF^\mathrm{PK}\cdot \hat\bb)\hat\bb = -f_0 (\hat\bb\cdot \hat\ee_x)\hat\bb, \label{smeq:Fglide} \\
    \FF^\mathrm{PK}_\mathrm{climb} &= (\FF^\mathrm{PK}\cdot \hat\bb^\perp)\hat\bb^\perp = -f_0 (\hat\bb^\perp\cdot \ee_x)\hat\bb^\perp.  \label{smeq:Fclimb}
\end{align}

\textbf{Glide.} Let us focus  on the glide component first (\eqref{smeq:Fglide}).
As seen in our numerical simulations, this is the main direction in which the dislocations move, Fig.~\ref{fig:disloc}C. 
The amplitude of the glide force is proportional to $|\cos\beta|$ where $\beta$ is the angle of the Burgers vector with the horizontal axis. 
More relevant is the speed at which a gliding dislocation moves in the $x$ direction. It is speed is proportional to $-\cos^2\beta$, in agreement with our numerical simulations, see Fig.~\ref{fig:disloc}F.

Instead of the effective Peach-Koehler argument, one may adopt a purely kinematic perspective on the glide of the dislocations as in the main text. 
Microscopically,  glide happens if particles above and below  the glide line are driven in opposite  directions (parallel the glide line). 
Noting $\dot{\mathbf u}^{\rm NR}$ the velocity imposed by the hydrodynamic forces  its component along the glide line is $\dot{\mathbf u}^{\rm NR}\cdot\hat\bb$. 
Writing the gradient perpendicular to the glide line as $\hat\bb^\perp\cdot\bm\nabla$, we conclude that the component of the velocity field $\dot \uu^{\rm NR}$ that powers motion is  $\hat\bb^\perp\cdot\bm\nabla(\dot \uu^{\rm NR}\cdot\hat\bb)$. 
This kinematic reasoning (which particles move in which direction) gives an equivalent perspective as the Peach-Koehler argument.

We note that at strong magnetic fields in our numerical simulations, gliding dislocations actually favor the ordering of the system since two of them can collide and annihilate or merge into a single one (see Fig.~3I of the main text, showing a circular grain boundary as initial condition).

\textbf{Climb.} The amplitude of the climb force, \eqref{smeq:Fclimb}, is proportional to $|\sin\beta|$. 
In practice, a dislocation cannot move in this direction because this would require a mass transfer between above and below the glide plane. An alternative scenario, clearly illustrated in Ref.~\cite{Irvine_2013}, was dubbed ``climbing by gliding''. 
It involves first the splitting of the dislocation into a pair of dislocations (with conservation of the total vector charge) and then the gliding of these two dislocations which have glide planes different from the one of the original dislocation. This scenario is observed in our numerical simulations, see Fig.~5B. It provides a basis for the splitting, and by extension the proliferation of dislocations  leading to the melting of the system when the magnetic interactions are weak enough.

\subsection{Grain boundaries}
The phenomenology that we observe for isolated dislocations in our numerical simulations also holds for grain boundaries. In particular, (straight) grain boundaries move upstream, see Fig.~3E of the main text and Fig.~\ref{fig:disloc}D-E. 
This is not surprising since  an interface between two cristals with a finite angle mismatch is nothing but a line of well separated dislocations (the value of the angle mismatch defines  the line density of the dislocation along the grain boundary).
These dislocations glide upstream together so that the grain boundary globally translates in this direction.
(We recall that the angle mismatch $\Delta\theta_6\in [-30\si{\degree}, 30\si{\degree}]$ is defined as the difference between the phases $\theta_6$ of the hexatic order parameter in the two phases).  

The dependence of the speed of a grain boundary, as a function of its orientation from the horizontal axis is similar to an isolated dislocation's, but with a smaller prefactor, see Fig.~\ref{fig:disloc}F.

\bibliography{biblio.bib}